# A user-centered approach to designing an experimental laboratory data platform

Ha-Kyung Kwon[a*], Chirranjeevi Balaji Gopal[a*], Jared Kirschner[b], Santiago Caicedo[b], and Brian D. Storey[a]

[a] Toyota Research Institute, Los Altos, CA 94022, [b] Continuum-EPAM, Boston, MA 02210

*These authors contributed equally

## Abstract

While automated experiments and high-throughput methods are becoming more mainstream in the age of data, empowering individual researchers to capture, collate, and contextualize their data faster and more reproducibly still remains a challenge in science. Despite the abundance of software products to help digitize and organize scientific information, their broader adoption in the scientific community has been hindered by the lack of a holistic understanding of the diverse needs of researchers and their experimental processes. In this work, we take a user-centered approach to understand what essential elements of design and functionality researchers (in chemical and materials science) want in an experimental data platform to address the problem of data capture in their experimental processes. We found that having the capability to contextualize rich, complex experimental datasets is the primary user requirement. We synthesize this and other key findings into design criteria for a potential solution.

## Keywords

## Introduction

It typically takes decades to go from the invention of a new material in the laboratory to its use in technology [1]. Because bringing new materials to market is critical to many sectors, there have been collective efforts over the past few decades focused on accelerating materials development. One recent effort, the Materials Genome Initiative, led to an explosion of interest in using tools from data science, machine learning, and artificial intelligence (AI) to accelerate materials discovery [2–5]. More recently, attention has expanded toward merging AI with robotics to create automated laboratories [4,6,7]. While the latter holds promise for the future, the reality is that a lot of scientific research will continue to be done in traditional, human-centered laboratories, synthesizing new materials or modifying existing ones in light of new research. A plausible future innovation lies in the development of a software platform that can integrate traditional data collected from human-centered laboratories with artificial intelligence and data from automated, high-throughput experiments [8].

Building a data platform for traditional laboratories brings many challenges, such as low sample count, process variations across labs, lack of adequate metadata, lack of dark data (i.e. results from failed experiments) and dynamic, non-standard workflows. Such a platform must enable a wide range of data and metadata capture from a multitude of experiments and be amenable to collaborations across a wide network. Aside from these challenges, the success of *any* platform relies on a growing community of users that can mutually benefit from it. For an experimental data platform, driving user adoption requires focusing on the *human* in human-centered labs. Driving adoption is challenging, since the intended benefit of accelerating the pace of discovery is hard to quantify, and if that benefit ever comes, it is unlikely to be attributed to the original researchers. For individual researchers to invest in such an experimental data platform, it should primarily make their work easier.

Electronic Lab Notebooks (ELN) are commonly deployed tools for systematic capture and storage of data in research environments. They offer advantages of reduced data loss associated with manual data handling, long-term storage, provenance tracking, workflow templates, and the ability to collaborate and retrieve records across multiple devices [9]. A multitude of ELNs are available in the market, which are often adapted to address domain-specific challenges, or new variants designed ground-up [10–13]. For instance, chemical structure drawings are integral to experimental planning in chemical sciences, and having this feature in an ELN can make or break its adoption [14]. To organize the unstructured scientific knowledge contained in ELNs, semantic web technologies have been proposed [15,16]. Despite the wide range of options in the ELN space, none has received widespread engagement in academic research, suggesting a lack of fit between what is offered and the needs of researchers. The recent analysis of available ELN tools by Kanza *et al.* revealed that cost, ease of use, and accessibility across devices are the main barriers to broader adoption [17]. Researchers want ELNs to assist, not replace, their lab notebook. The need for a user-centric design to improve adoption has been recognized in prior work, but the efforts have been within the constraints of an ELN-based solution [16–19].

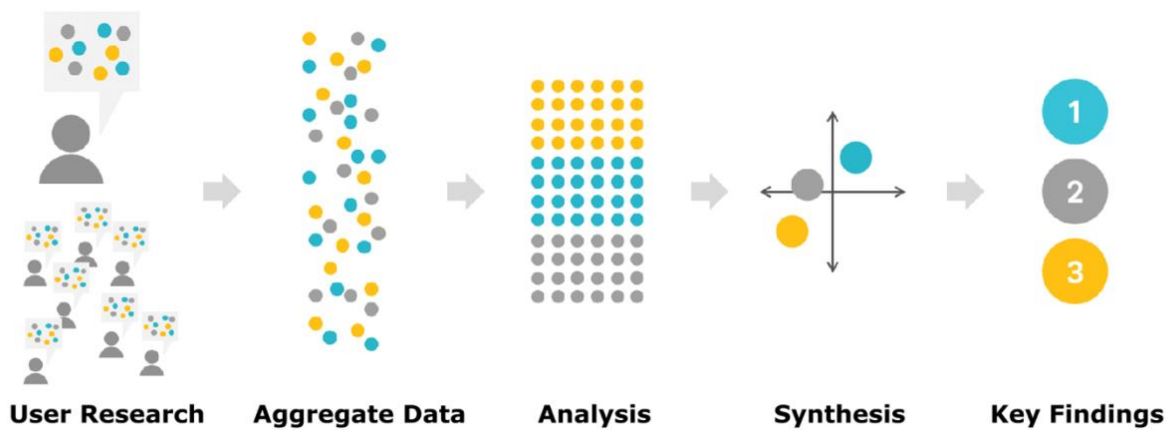

Figure 1: Illustrative summary of the user-centered design approach to problem framing. We talk to cohorts of users to understand their pain points, aggregate the data, analyze them for patterns and commonalities, and then synthesize them into key findings to which a potential product can be tailored.

In this manuscript, we describe our efforts to better understand user pain points with data management in research workflows and put forth design requirements for a potential solution. Our primary motivation was to take a user-centered design approach to *frame the problem space* – unconstrained by solution– in the words of researchers. This framing would serve as a design guide for an experimental data platform that researchers across diverse scientific domains would willingly use in their daily activities, since it streamlines data management and the process of scientific inquiry based on experimental laboratory research. We conducted ethnographic research with 15 participants from three research groups and two universities working in chemical engineering and catalysis, polymer synthesis, and electrochemical devices and materials characterization. None of the researchers we spoke with actively use an ELN because ELNs try to replace the simple, yet effective, lab notebook. We propose that rather than digitizing laboratory notes, an effective experimental data platform should prioritize assisting researchers in comparing and analyzing the data they accumulate and enabling creation of new scientific hypotheses or arguments from it.

**Methodology**

We employed human-centered design philosophy [20,21] to derive key product insights and goals from raw user research data. A schematic of the process is shown in Figure 1. The methodology relies on in-depth interviews and direct observations of a small but representative group of people, as opposed to broad surveys which collect limited data from a larger group. Another key aspect is to base conclusions on observations of how people behave, as opposed to what they may say in an interview setting only.

| Experience | | Discipline | | Expertise | |
|---|---|---|---|---|---|
| Undergraduate | 1 | Chemistry | 7 | Synthesis | 7 |
| PhD student | 5 | Chemical Engineering | 6 | Characterization | 6 |
| Postdoctoral | 7 | Physics | 1 | Advisory | 2 |
| Principal Investigator | 2 | Materials Science and Engineering | 1 | | |

Table 1: Distribution of experience, discipline, and expertise in our users. A total of 15 users across three academic labs in the United States were interviewed. Discipline of each user was determined by the department that awarded the user's most recent degree.

Fifteen researchers from three academic research groups were recruited in order to ensure that a diversity of experiences, disciplines, and expertise were represented, as shown in Table 1. Further details regarding the selection criteria can be found in Supplementary Table 1. Each researcher was interviewed and subsequently observed in the process of recording and analyzing information inside and outside of labs. Following user interviews, raw research data (e.g. contemporaneous notes, transcripts, recordings, pictures) from each interview were aggregated and distilled into a summarized form, Figure 2(a). This distilled information was then analyzed for recurring patterns, Figure 2(b). Important observations and direct quotes were grouped into relevant clusters, Figure 2(c). User needs underlying the patterns were synthesized into *Key Findings* that inform the right problems to solve based on key researcher pain points. These findings then lead us to a set of *Design Principles* which are solution-agnostic and should be followed by any proposed solution. We recognize that the sample-size of fifteen might be somewhat limited, but it allowed us to perform a deep-dive with every individual and benchmark the methodology to be used on a larger corpus of interviewees in chemistry and chemical sciences. More details on the interview process and subsequent analysis are provided in the Supplementary Materials.

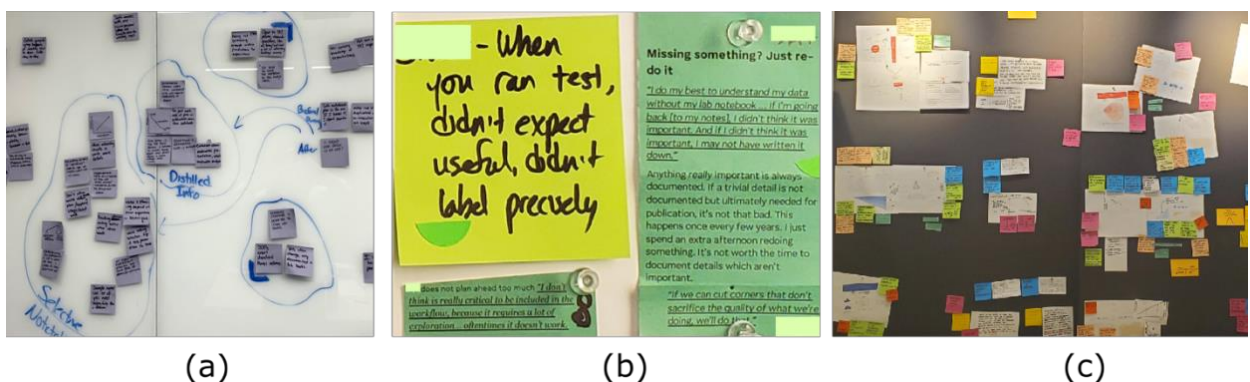

Figure 2: (a) Distillation of raw notes from a user interview. We work with summaries from each interview to analyze patterns across multiple users. (b) A specific group of related observations. This specific group of data contributes to both "Key Finding 6: Not Enough Breadcrumbs" and "Design Principle: Focus on Results ". (c) Macro view of (a), showing multiple clusters of related direct quotes, observations, and interpretations. Clusters are iteratively updated as we continue to consider the user research data and whether the cluster suggests something insightful regarding the problem or solution.

## Results and Discussion

**Key Findings**

In this section, we describe our eight key findings (KF) from user research. Key findings reflect the needs of a user and are hence written from the perspective of a researcher.

**KF1: Manuscripts aren't everything**

*"This material is not new in the literature, so it was really frustrating for me that no one had mentioned that it wasn't stable. I [presented] posters on it and got some really delightful feedback from a professor like, 'Why did you keep going? It's clear that this [material] isn't stable."*

The research process often relies on existing literature, applying what is useful to the current research question. Unfortunately, manuscripts are often missing helpful information, such as specifications of experimental setup, exactly how the measurements were taken, what paths did not work, etc. While such details are appended after the main text, or in supplementary materials, critical information is still missing due to manuscript space limitations, traditional expectations of what is publication-worthy, or simply because it is unknown to the author.

This issue is compounded by the paradigm of publication-as-a-currency in promoting one's work. Publishing high-impact work requires research findings that are new, interesting, and successful. Researchers focus on documenting their successes because failures will not help them get recognition. Though there may be scientific value in the failed (dark) data, it will likely be found by others after they have moved on, so why take that risk with their limited time now? Ultimately, new researchers waste time rediscovering known information, and the original researcher does not benefit from external reproduction attempts.

**KF2: No recordkeeping standards**

*"If my [file] naming scheme isn't very understandable to someone I'm working with, it's probably not understandable to anyone else."*

The principal investigator is often removed from the day-to-day operations of the lab and mostly focuses on results and scientific directions. Researchers are trusted to decide what recordkeeping system works best for them. They adapt practices from peers and mentors, but there is often enough difference between approaches that one cannot understand another's system. Some common threads in the practices included the incorporation of metadata, such as the date, material name, composition, and researcher name, into the name of a data file and the use of visual cues to highlight important facts within data tables and observations.

The lack of standardization is not a problem for a researcher who works individually and on a single task at a time. Problems arise when a former lab member's work would help a current researcher, but they are no longer around to provide guidance. Parsing another's idiosyncratic recordkeeping system is possible, but tedious.

Interestingly, the only time we observed somewhat standardized recordkeeping practices is in collaborative experiments, where the detail and quality of recordkeeping is dictated by the need to work together. Researchers take measures to describe, explain, and share internalized knowledge, so that the records can be shared without room for interpretation or confusion.

This particular *key finding* highlights an opportunity for the principal investigator or mentor to encourage a recordkeeping standard, especially if parts of it are already being adapted by the researchers. Kanza *et al* found that social influence–by a mentor or an adviser– can be an important factor in driving adoption of new tools[22].

### KF3: Need version control
*"[The computational team] did maybe 3 or 4 sets of calculations [each in different Excel sheets]. I used those parameters to make a plot. I think later they updated the file, but my plot wasn't updated."*

Though researchers often lead their own projects and work independently, most projects cannot be completed alone due to the range of specializations involved. When collaborating on a project, data tends to be managed at the individual level rather than in shared, mutually understood locations and formats. The project lead requires high-level, summary information from the specialists, while the underlying details remain in individual spreadsheets and files.

Researchers often interpret someone else's spreadsheet and copy into their own, differently formatted spreadsheet. Occasionally, someone will update information without a dependent person realizing it. While long-time collaborators have established methods of recording and sharing information, finding and agreeing on these methods of working together can be a prolonged process for new collaborators.

### KF4: What I need evolves over time
*"The flexibility of the plan is important. Everything goes out the window when you start [executing a plan], because nothing ever works."*

Many current ELNs for data logging come with prescriptive templates, which are ideal for mature workflows. In contrast, academic research is a highly cyclical process, involving many iterations of planning, execution, analysis, and subsequent modification. Researchers generally document what they did, less so what they are planning to do. Mistakes are inevitable and plans evolve constantly, so entering information beforehand feels futile. There is a tendency to prioritize the immediate search for a promising direction over detailed documentation, as a lot of ideas seldom work the first time. In addition, the definition of a "promising direction" changes depending on new findings, which means that the needs of the researcher– and the types of data and observation that need to be captured– often change over time. The constantly evolving plans, coupled with the fear of being judged (discussed in detail in KF7) for mistakes or failures, make researchers less likely to document and share work in progress.

**KF5: Needle in information haystack**
*"I know the data is there, but [...] I don't know where it is, though I'm 100% sure I have [it]."*

Data is often generated from diverse instruments and stored in a variety of formats, in different locations over long periods of time. Researchers try to maintain a consistent system of organization and cross-references across digital platforms (data files, documents, spreadsheets) and physical locations (samples, handwritten notes). They also try to keep important information (such as promising sample information, data, and metadata) in one place as they progress, such as in filenames or presentations.

However, when they need to plan, analyze data, or create a presentation, there is often something they know exists but cannot easily find. Even when much of their information is digital (i.e. stored in laptops or cloud storage), we observed that researchers require minutes to find specific information during our interviews. Their system makes sense to them with some challenges, but it is nearly impossible for others to navigate.

**KF6: Not enough breadcrumbs**
*"We're all working on several different projects, and we don't really have a system right now that keeps track of [...] the last experiment we've run on a project. You just have to remember, or hopefully you have a research update."*

Researchers often need to review their work to understand the current status of a project, for instance, when switching between multiple projects, preparing a presentation or a manuscript, or revisiting old work with new inspiration.

Going back to prior work is challenging because research is almost never straightforward, and the direction of a project may change based on new findings and the researcher's updated knowledge. Researchers navigating the process are focused on the path of least resistance to interesting, successful results, rather than mapping every step of every investigation. Recreating the entirety of a researcher's journey beyond just the final, successful path requires access to their thought process in addition to the accompanying data.

This surprising finding is critical to understanding many of researchers' behavior. In the process, recorded data are not curated with metadata or details, because the context is obvious to the researcher actively working on the project. It is only when a future collaborator needs to understand the decision-making behind the data that this lack of breadcrumbs poses a challenge.

**KF7: Sharing is vulnerability**
*"Nobody wants to hear a presentation full of 'I tried all of these things, they didn't work.'"*

Researchers feel social pressure against sharing unsuccessful outcomes and unpolished material. This barrier increases as the audience becomes larger, more formal, and less intimate, and can prevent the sharing of important material such as dark data (what does not work), raw data (prior to processing and baseline corrections), and observations that contain personal feelings of frustration or triumph.

Lab notebooks are implicitly understood to be a temporary and personal log of one's experimental journey, consistent with findings of Kanza [22]. They capture both successes and failures with just enough detail for the specific researcher to understand, and often no one else. The content is of value while in the lab and any long-term information is copied to other secure forms. The lab notebook is almost never shared without the owner present. Requesting raw data and notes can sometimes imply mistrust and disbelief in the researcher's analysis and experiments.

**KF8: Presentation goldilocks zone**
*"I go back to PowerPoints a lot [because] those are the most distilled [versions], other than the papers, of what I do. If I want to remember what I was doing on this project, what the data looked like, I will go back to the PowerPoints first."*

Presentations are the most important in-process artifact for a researcher. Creating a presentation allows a researcher to review the project from a higher level, and curate and contextualize research progress. Complete presentations containing a lot of results are often created for meetings within a group.

When researchers revisit past work, presentations are the first point of reference. But striking the balance between what is valuable to a researcher and to their audience is difficult. It takes much more effort to make something understandable and polished enough for their various audiences than for just themselves. Researchers also exclude details that matter to them but not to the audience, making the presentation an imperfect reference.

**Design Principles**
The key findings inform our understanding of the pain points and contexts for experimental researchers. Based on the key findings, we have developed design principles to guide the development of an experimental data platform. They act as a design compass, providing constraints and guidelines to any proposed solution to ensure consistency in the user experience.

**DP1: Focus on results**
In order to achieve widespread adoption, the platform must prioritize what is important to users: getting to result quickly by organizing, visualizing, and analyzing data. While processes, trials, and know-how are important to document, prioritizing results can quickly lead to publications, bringing immediate value to researchers (**KF1**). As we saw in **KF6**, researchers are focused on mapping out a successful path from ideation to conclusion. In order to immediately address these needs, the platform must provide a way to contextualize insights and technical details, in addition to collecting data and organizing key results. Collecting appropriate context such as how, when, and why the data was collected, the environment it was collected in, and any relevant metadata can significantly help users make subsequent decisions and troubleshoot results later.

**DP2: Show me the way**
Keeping a bird's eye view of all projects that are being worked on can help the researcher review their work with a fresh insight. A researcher has no time to revisit all their paths. Providing a simple and easy way to view the relevant goals for each project can decrease overhead and help a researcher pick up their work more quickly (**KF6**), and to find related experimental notes and data more efficiently (**KF5**). Such a bird's eye view can also enable researchers to see when previous work might be relevant for a current project.

**DP3: I control who sees what**
Protection and security are a requisite for sharing (**KF7**). The level of sharing and openness depends on the level of collaboration and trust. The platform must allow a user to decide what data –whether it be digital, or physical– can be shared and safely discussed and with whom in the scientific community. The audience evolves and broadens as projects mature.

**DP4: Enhance but don't replace paper**
A researcher's notes and notetaking practices evolve over time as the experimental techniques and methods change (**KF4**). Paper provides a facile, efficient, and flexible way to write down notes in the lab, and to bring in a synthesis recipe or a plan and later append to the lab notebook. This type of notetaking is especially important since researchers prefer to keep things inside the lab separate from those outside for concerns of contamination and safety. They understand that handwritten notes cannot be easily searched and thus cannot be used as the only method of notetaking; however, a digital platform cannot easily replicate the feel and ease of use of paper.

**DP5: Coexist with my tools and ways of working**
The platform must accommodate the different preferred tools and methods with which individuals organize information, analyze scientific data, and formulate their thoughts (**KF2**). It must also allow different specializations to collaborate and share information (**KF3**). Out of necessity, a researcher stores information in both computers and lab notebooks (**KF5**). The platform must reduce the effort on the part of the researcher to integrate the diverse data streams into it. In addition, researchers may use the same tool for multiple purposes: some use presentation slides to share high-level information with peers, while others use detailed slides as a way-finder when revisiting their own work (**KF8**).

An experimental data platform must integrate not only existing tools, but also the ways in which they are used. The researcher's need to keep devices used in the lab separate from those outside means the experimental data platform must be accessible from devices in both locations.

**DP6: Credit me whenever possible**
The experimental data platform must encourage sharing of knowledge by crediting researchers for all their contributions (successful and failed experiments, knowhow, insights). In order to do so, researchers must be protected from others taking unfair credit for their work (**KF7**). Crediting one's contributions beyond published results can also help promote researchers and can form the basis of a community where unpublished results and processes can be shared for the benefit of all (**KF1**). The current tools for communicating with other researchers –through manuscripts, sharing of files, and presenting findings to others– often limit the usefulness of shared information between individuals and groups.

**Future work: from problem framing to solution**

Based on our user research, we are developing a software prototype with an emphasis on data contextualization. A conceptual overview of the problems that the software platform will address and the core components is provided in this section, while saving specifics of design, implementation details, and user-testing for a future publication. There are two components, which we label as *Evidence* and *Arguments* for ease of reference.

*Evidence* enhances the way researchers organize and search their data by aggregating and connecting all synthesis and characterization information in one place with a user-specified unit of organization, such as by sample or technique. *Evidence* presents data to the user along with the context in which it was collected, facilitating easier interpretation by the user and knowledge transfer to collaborators.

*Arguments* are built upon analysis and integration of evidence and findings. By viewing a set of *Arguments*, a researcher can quickly understand what they have tried, what has worked or not, what they have left to answer, and how close they are to answering their research question. *Arguments* are connected to their supporting *Evidence,* enabling users to easily find relevant information without having to search through many files and locations. Researchers can use *Arguments* to create scientific narratives.

*Evidence* and *Arguments* can be understood from the perspective of the researcher's experience in a chemical/materials science lab, demonstrated by a series of stories in Figure 3**.** Researchers often manage multiple directions of work to continually make progress even if a specific direction hits a road-block. This context switching comes with a significant cognitive load. The *Arguments* feature allows the researcher to quickly view progress on specific research questions, supporting evidence, and outstanding questions to address from previous analyses and presentations (Figure 3(a)). When researchers synthesize and characterize a new sample in the lab, *Evidence* allows inclusion of diverse data streams such as observational notes (images of plans written on loose paper and lab notebooks), quantitative data (from structural characterization) and environmental data (recorded as notes, or through direct upload from sensors) and organizes them all based on the sample (Figure 3(b)). In Figure 3(c), we describe a situation where one researcher must work with another to perform a specific characterization of the sample. The researcher performing the characterization can use *Evidence* to view the synthesis history of the sample, and then upload the results after characterization along with relevant documentation and raw instrument files. *Evidence* associates all measurements with relevant instrument details to the right sample identifier, ready for collaborating researchers to review. The *Arguments* feature allows researchers to compare multiple measurements and integrate evidence they have collected in order to build arguments toward supporting or refuting a hypothesis (Figure 3(d)). The grouping of evidence can be based on samples, techniques, or any other user-input variable. *Arguments* enables the user to expand their bandwidth for making persuasive arguments with the data they have collected.

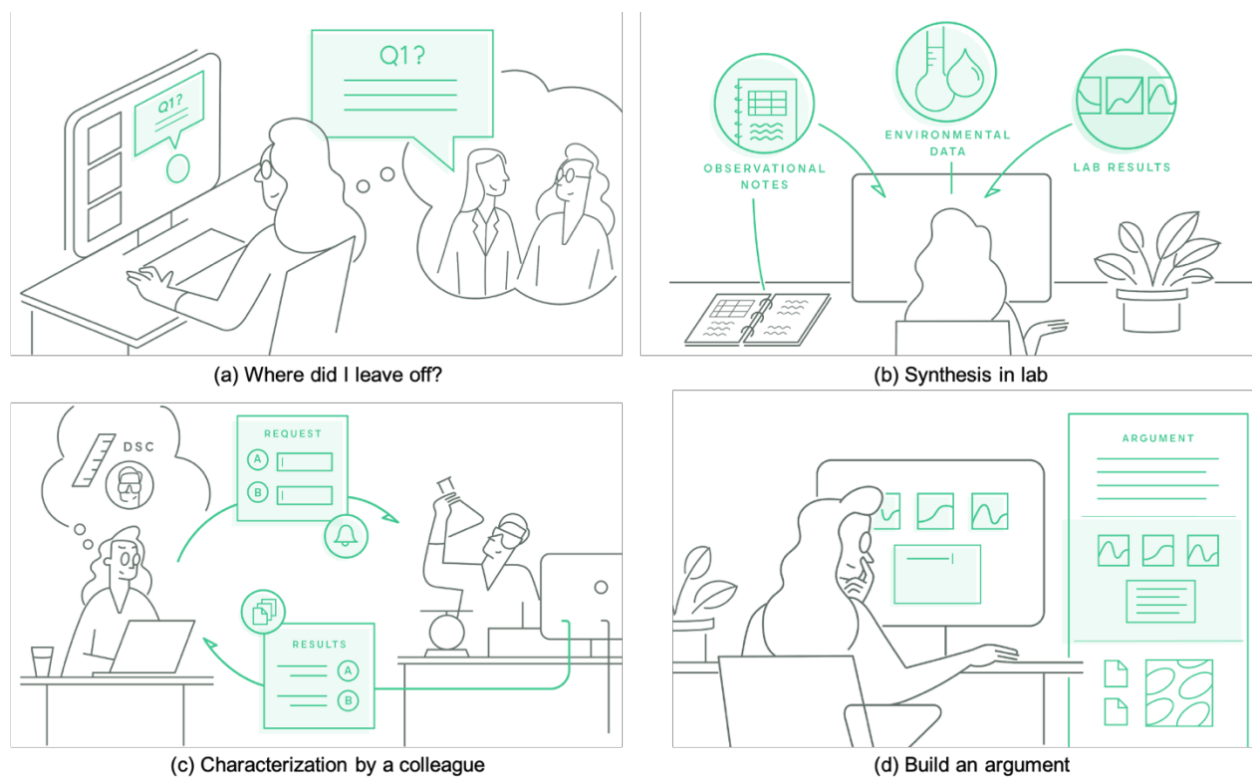

Figure 3. An illustration of some of the desired capabilities in a software in the form of user stories. The software allows users to organize all their synthesis and characterization data in one place, effectively collaborate with others, and integrate evidence to build scientific arguments.

**Conclusions**

In this work we took a user-centered design methodology to frame the requirements of an experimental data platform to help accelerate knowledge creation and integrate data from human-centered labs. To get a thorough understanding of the needs of an experimental researcher and prioritize them at the core of a potential solution, we conducted user research with researchers of varying experience, spanning three academic groups in chemical and materials sciences. Based on this, we propose the key enabler to be *contextualization of rich, complex experimental data*, beyond serving as a platform for digitizing notes and organizing data.

This challenge of data contextualization is not unique to the experimental domain. Even in the virtual world, where, by its very nature, data is presumed to be digitized, understanding the context and decision-making around data remains a challenge. Multiple systems have been developed to capture and utilize provenance in high-performance computing simulations [23]. Similarly, for machine learning experiments, open-source tools [24, 25] have been developed to easily track the provenance and sensitivity of hyperparameters. Contextualization of data is extremely valuable for a researcher looking to reproduce somebody else's work, or revisiting their own work.

However, because this information is highly internalized and frequently evolving, it is rarely documented in-process.

Our study suggest that an experimental data platform that aids researchers in aggregating and contextualizing different pieces of evidence (synthesis details and observations, characterization data and metadata, for instance) and enables them to construct a scientific narrative would be of immense value. The solution could eventually be extended to capture all notes, data, and metadata, along with provenance. Our hope is that the methodology used in this study will be tested on a broader corpus of interviewees to help generalize the findings and design principles, which subsequently serve as a foundation for a future platform that evolves with community participation from other experimental labs and developers.

**Declarations**

**Availability of data and materials**
The data supporting the results and conclusions of this article are included as supporting information.

**Competing interests**
The authors have no competing interests. All university research subject participants are funded in whole or in part through Toyota Research Institute, Los Altos, CA 94022.

**Funding**
The work was funded entirely by Toyota Research Institute, Los Altos, CA 94022.

**Authors' contributions**
All the authors conceived and planned the study. JK and SC designed the user-research study. All authors participated in conducting user interviews and analysing the resulting data. HK, CG and BS wrote the manuscript.

**Acknowledgements**
We are grateful to the Accelerated Materials Design and Discovery group at TRI for insightful discussions during the data collection and analysis phase. We would specifically like to thank Muratahan Aykol, Linda Hung, Joseph Montoya and Jens Hummelshoej for review and feedback on the manuscript.

**Notes and references**

1. Eagar TW (1995) Bringing new materials to market. Technol Rev 98:42–49

2. Reyes KG, Maruyama B (2019) The machine learning revolution in materials? MRS Bull 44:530–537

3. Gomes CP, Selman B, Gregoire JM (2019) Artificial intelligence for materials discovery. MRS Bull 44:538–544

4. de Pablo JJ, Jackson NE, Webb MA, et al (2019) New frontiers for the materials genome initiative. npj Computational Materials 5:41

5. Tabor DP, Roch LM, Saikin SK, et al (2018) Accelerating the discovery of materials for clean energy in the era of smart automation. Nature Reviews Materials 3:5–20

6. Service RF (2019) AIs direct search for materials breakthroughs. Science 366:1295–1296

7. Pendleton IM, Cattabriga G, Li Z, et al (2019) Experiment Specification, Capture and Laboratory Automation Technology (ESCALATE): a software pipeline for automated chemical experimentation and data management. MRS Communications 9:846–859

8. Aykol M, Hummelshøj JS, Anapolsky A, et al (2019) The Materials Research Platform: Defining the Requirements from User Stories. Matter 1:1433–1438

9. Bird CL, Willoughby C, Frey JG (2013) Laboratory notebooks in the digital era: the role of ELNs in record keeping for chemistry and other sciences. Chem Soc Rev 42:8157–8175

10. Bromfield Lee D (2018) Implementation and Student Perceptions on Google Docs as an Electronic Laboratory Notebook in Organic Chemistry. J Chem Educ 95:1102–1111

11. Mohd Zaki Z, Dew PM, Lau LMS, et al (2013) Architecture design of a user-orientated electronic laboratory notebook: A case study within an atmospheric chemistry community. Future Gener Comput Syst 29:2182–2196

12. Walsh E, Cho I (2013) Using Evernote as an electronic lab notebook in a translational science laboratory. J Lab Autom 18:229–234

13. Riley EM, Hattaway HZ, Felse PA (2017) Implementation and use of cloud-based electronic lab notebook in a bioprocess engineering teaching laboratory. J Biol Eng 11:40

14. Tremouilhac P, Nguyen A, Huang Y-C, et al (2017) Chemotion ELN: an Open Source electronic lab notebook for chemists in academia. J Cheminform 9:54

15. Coles SJ, Frey JG, Bird CL, et al (2013) First steps towards semantic descriptions of electronic laboratory notebook records. J Cheminform 5:52

16. Kanza S, Gibbins N, Frey JG (2019) Too many tags spoil the metadata: investigating the knowledge management of scientific research with semantic web technologies. J Cheminform 11:23

17. Kanza S, Willoughby C, Gibbins N, et al (2017) Electronic lab notebooks: can they replace paper? J Cheminform 9:31

18. Reimer YJ, Douglas SA (2004) Ethnography, Scenario-Based Observational Usability Study, and Other Reviews Inform the Design of a Web-Based E-Notebook. International Journal of Human–Computer Interaction 17:403–426

19. Badiola KA, Bird C, Brocklesby WS, et al (2015) Experiences with a researcher-centric ELN. Chem Sci 6:1614–1629

20. Norman DA, Draper SW (1986) User centered system design: New perspectives on human-computer interaction. CRC Press

21. Abras C, Maloney-Krichmar D, Preece J, Others (2004) User-centered design. Bainbridge, W Encyclopedia of Human-Computer Interaction Thousand Oaks: Sage Publications 37:445–456

22. Kanza S (2018) What influence would a cloud based semantic laboratory notebook have on the digitisation and management of scientific research? Phd, University of Southampton

23. Suh Y-K, Lee KY (2018) A survey of simulation provenance systems: modeling, capturing, querying, visualization, and advanced utilization. Human-centric Computing and Information Sciences 8:27

24. Biewald L (2020) Experiment Tracking with Weights and Biases

25. Zaharia M, Chen A, Davidson A, et al (2018) Accelerating the Machine Learning Lifecycle with MLflow. IEEE Data Eng Bull 41:39–45

**A user-centered approach to designing an experimental laboratory data platform**

Ha-Kyung Kwon[a*], Chirranjeevi Balaji Gopal[a*], Jared Kirschner[b], Santiago Caicedo[b], and Brian D. Storey[a]

[a] Toyota Research Institute, Los Altos, CA 94022, [b] Continuum-EPAM, Boston, MA 02210

*These authors contributed equally

**Supplementary Materials**

**Cohort Selection**

We conducted ethnographic studies with 15 researchers, chosen from three academic research groups with active research funded by Toyota Research Institute, though participation in the study was purely voluntary. The participants were chosen in order to achieve balance in the recruiting criteria listed in Supplementary Table 1, ensuring that multiple levels of experience, discipline, and expertise were represented, as listed in Supplementary Table 1.

| Category | Type |
| --- | --- |
| Role | Synthesis workflow/researcher |
| | Characterization workflow/researcher |
| | Research lead/data accessor |
| Tool usage | Has used an ELN in the past year |
| | Has never used an ELN |
| Collaboration | Has worked with non-collocated collaborators |
| Time horizon | Hard end-date with group in < 2 years |
| | No hard end-date |
| Tenure | Has been in the group for < 1 year |
| | Has been in the group for > 1 year |
| Co-design | Is interested in continued project involvement and feedback |

Supplementary Table 1: Recruiting criteria to ensure the diversity of relevant perspectives was represented in the interviewee pool

**User research and data aggregation**

Each participant was interviewed according to a discussion guide we developed in order to understand researcher needs and behavior patterns in experimental laboratories.

The discussion guide for researchers was deliberately segmented to probe three key behaviors.

- Current record keeping and pain points: How and when is information recorded, referenced, stored, and shared?
- Nature of collaborative projects: At what points does collaboration occur in the lifetime of a project, and what do they look like with collaborators on-site and remote?
- Change: What are the drivers and inhibitors of behavior change in the lab, and how can we incentivize adoption of new processes and tools?

Rather than serving as a strict interview script, the discussion guide was designed to help us explore the motivations of researchers for doing what they do in and out of the lab. The interview process is designed to understand not just what the researcher thinks or does, but why.

Participants were also asked to walk us through how they use their current record keeping tools (lab notebooks, pieces of paper, and computers), so we could learn their process of storing and finding information by observation and understand their mental models for organizing information. While direct observation of the work environment is preferred over interviews, given that many of the lab processes can take months to complete from end to end we employed a combination of interviews and observations.

In addition to targeted interview questions, our user research contained three activities to facilitate conversations about the mental processes and priorities of participants, as shown in Supplementary Figure 1.

- Research process diagram, in which participants were presented with the labels "Explore & Plan", "Execute", "Analyze", and "Report" and asked to illustrate their research process at a high level.
- Feature rank, in which participants were presented with 12 cards containing attributes related to information and knowledge management and asked to prioritize and explain their importance. Specifically, they were asked to focus on 1-3 attributes most and least important to them. The goal is not to quantify responses, but instead to understand why.
- Information access map, in which participants were asked to discuss who should have access to different types of their research related information and why.

Researcher interviews were followed by an observation of each group of participants in their labs to fully understand the context of their working environment and any physical constraints, such as the layout of the laboratory, available benchtop space, and in-lab use of computers or notebooks.

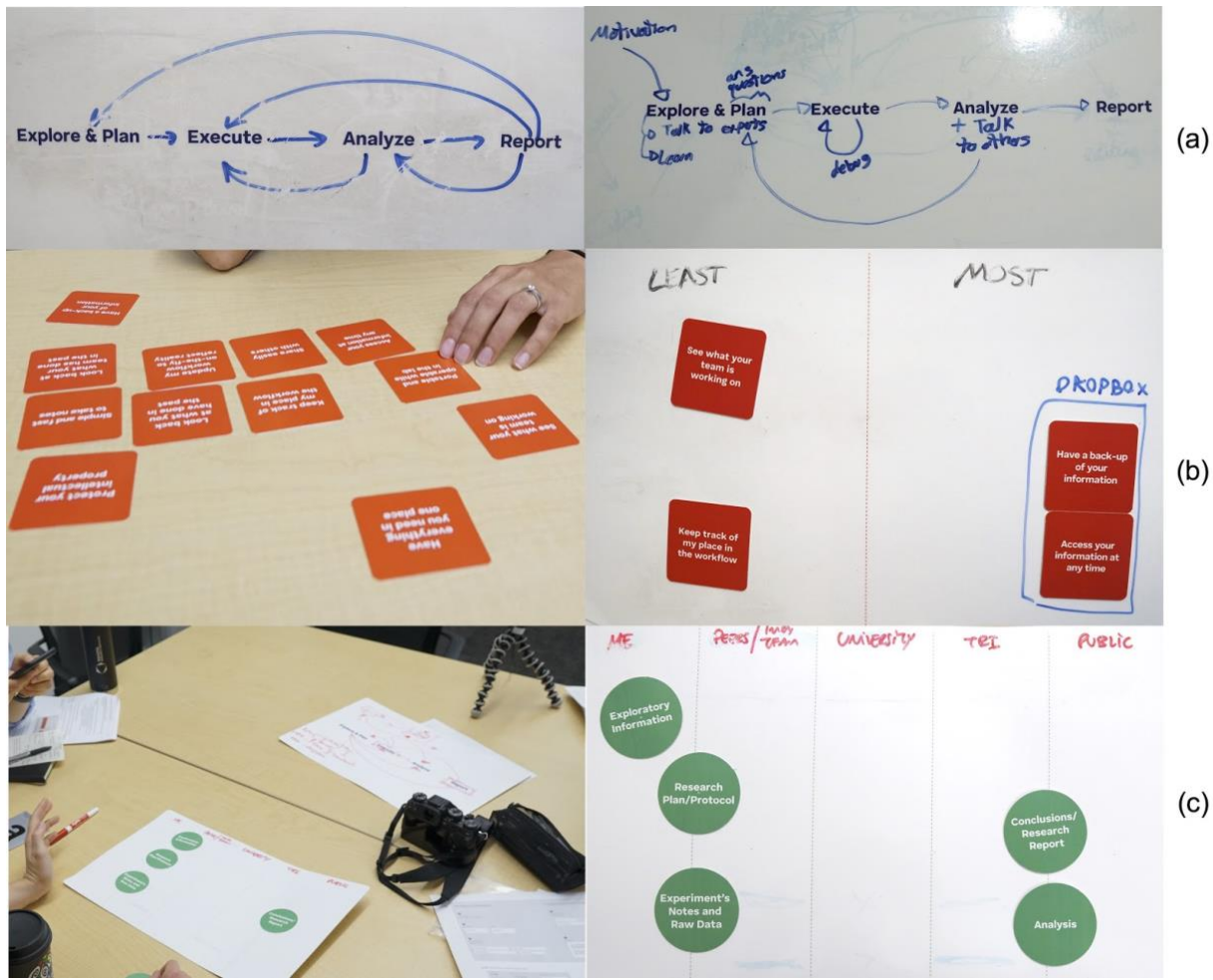

Supplementary Figure 1: (a) Responses from two researchers for the research process diagram activity. This activity revealed how iterative the research process is, and that there are many intermediate forms of reporting (e.g., meetings with experts, monthly reports to the PI, presentations), not just manuscripts. (b) Snapshot of the feature rank activity for a user indicating that "See what your team is working on" was the least important. For said researcher, that is because they already learn what they need to about the work of lab mates from lab meetings and don't want to be bombarded with additional files and information not relevant to their own work. (c) A sample information access map that helped us understand what researchers wanted to keep private, *and* why—though researchers wanted to share polished, final results broadly to advance human knowledge, they were hesitant to share work in-progress due to concerns of being judged for ideas that didn't work out, mistakes, fear of being scooped and the emotional/informal nature of lab notes.

**A user-centered approach to designing an experimental laboratory data platform**

Ha-Kyung Kwon[a]*, Chirranjeevi Balaji Gopal[a]*, Jared Kirschner[b], Santiago Caicedo[b], and Brian D. Storey[a]

[a] Toyota Research Institute, Los Altos, CA 94022, [b]Continuum-EPAM, Boston, MA 02210
*These authors contributed equally

Supplementary Materials - Researcher Discussion Guide

# Researcher Discussion Guide

**Project Eidetic**

**120 Min**

**Goal:** For TRI to capture today's experimental data from collaborators, as a stepping stone towards deriving value from that data to accelerate research and development. The main goals of this effort are to understand collaborating researchers' perspectives on:

- **Recordkeeping:** Understand current lab notebook behaviors and pain points: how information is recorded, referenced, stored, and shared; how this relates to workflows and their evolution
- **Collaboration:** Understand when and how collaboration occurs during an experiment, including with remote collaborators
- **Behavior Change:** Understand drivers and inhibitors of behavior change in the lab, such as for new tool or process adoption

**Team Roles:**

1) Primary interviewer (EPAM Continuum): Leads the conversation, handles the stimulus, captures key photos noted in the guide.
2) Lead note taker (EPAM Continuum): Start/stop audio recorder. Take notes throughout the interview, capturing feedback in the participants own words as much as possible. This ideally reads almost like a transcript (no need to capture exact wording of questions, just answers). Secondary role is to remind primary interviewer if they skipped a section, and to ask additional questions when invited by the interviewer.
3-4) Support (1-2 TRI): Capture additional relevant photos, note important observations on pen and paper, and ask additional questions when invited by the interviewer.

*Quick etiquette reminders: Arrive 15 min early. Dress casual and lab-appropriate. Turn phone to mute (and leave it in your pocket during the interview, unless being used to photograph). Smile and nod, with open body language – stay engaged and accepting, whether feedback is positive or negative.*

**Key photos:**
- of laboratory locations where notes are taken
- of notebook entries for TRI research
- of activities
- headshot of researcher

Note: This is a discussion guide. It is intended as a suggested flow of questions, and not a formal, set, script. While the major topics will be covered in each interview, the exact questions asked, and the sequence of the interview may vary depending on each individual interview.

Note: We follow a "perception is reality" philosophy. During the course of an interview, the facilitator will adopt the language of the respondent, even if it is not factually correct. In other words, if the respondent uses incorrect terminology, we will not correct them. Any discrepancies between respondents' perception versus factual reality will be captured during the debrief session and synthesized during the analysis of the project.

# Introduction [5 Min, Total 5 Min]
*Goal: Set goals and expectations.*

Thanks for taking time to talk with us. We really appreciate it. My name is _____, this is ____, and we're from EPAM Continuum, a design and innovation company. We've created products like the Swiffer by talking to people like you and designing around your needs. We're partnering with Toyota Research Institute – TRI – to improve the scientific research process together, which is why _____ and _____ from TRI are here with us today.

We're here to talk about your role as a researcher, how and when you collaborate with others in the lab, and how you track and share information throughout the research process.

You are the expert and we're here to learn from you! We are not scientists, so we may ask some naïve questions—please humor them!

*TRI team member:*
> There are no wrong or right answers, so please be candid and honest; your insight and perspective will help us make better solutions for researchers like you.

We want to be respectful of your time today. We are expecting to spend about 2 hours with you – is that still okay?

Before we begin, there are a few details I want to go over with you.
- We have a release form for you to review and sign. It states that you're getting compensation in appreciation for your participation, not to discuss anything we talk about today, and that we can use any ideas we get from this conversation. *[Wait for signature]*
- Here's a copy of the release form for you to keep, your gift card, and my business card in case you have questions or further thoughts later.
- We're going to take notes and photos, and record audio and video, for our internal note-taking purposes only. Nothing will be shared with your PI or other members of your group. Is that okay?
- I may look at my watch or redirect you today. This is only to make sure that I get your valuable perspective on everything we planned!

Do you have any questions for us before we begin?

# RESEARCHER'S CONTEXT [20 Min, Total 25 Min]
*Goal: Learn details about the participant and their 'natural habit', so we can understand their responses in context.*

1. Let's start with the basics. Can you tell me a bit about yourself?
2. Why materials science? Where did your interest start?
3. Can you describe what you do in your research group, your specific role and responsibilities?
    - Why did you choose this research group?
    - In what ways has it been different than what you were expecting?
    - Within your research group, what is deemed worthy of recognition by others? *(peers, PI, external)*
    - What does it mean to be successful in the group?
        - Is performance assessed in any way?
        - How does good performance or recognition affect your career?
4. Next, we'd like to explore the professional relationships and interactions that are important to you and your work. These can be with people, groups, institutions, regulations. Can you tell me about them?
    - *Probe other options if picture seems incomplete: PI, advisor, peers outside group, external collaborators, research sponsors, university, regulations*
5. What is your relationship with TRI? What interactions do you have with them?
    - What is the purpose of the interactions? When do they occur?
    - What expectations do you think TRI has? Does this influence anything about your process?
    - How does TRI enable your research?
6. Help me understand the relationship between words like project, experiment, and workflow.
7. At a high-level, we understand that these steps are involved in the process, from the very beginning to the very end. Can you trace us through the path by which things happen? What's linear? Cyclical? And is there anything big we've missed?
    - *Moderator: show process map, gesture with dry erase marker.*

    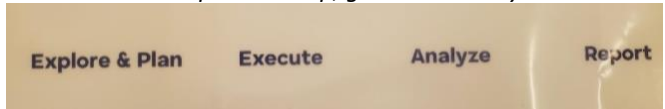

    - Are you not very involved in any of these stages?
        - *Response will be used to decide where to skip, if needed, in Process Deep Dive*
    - How many things are you working on at a time? Is there overlap?
        - What are the challenges?
    - Which stage is most important to you, and why?
        - What are some challenges that make your work more difficult at that stage?

# Process Deep Dive [60 Min, Total 85 Min]
*Goal: understand current recordkeeping behavior and pain points throughout experiment lifecycle: how information is recorded, referenced, stored, and shared.*

*Go through "Plan" and "Execute" for everyone, but skip "Explore", "Analyze", or "Report" at your discretion informed by what stage(s) are most relevant to them.*

Now that we've talked about your experimental process at a high level, we want to dig into some of the details. A part of this is seeing the ways in which you generate, record, and share information and knowledge along the way. During this part of the conversation, you'll need your recordkeeping system from your most recent Toyota Research Institute related experiment in whatever forms it exists. Do you have it with you?

As we go, we may ask you to show us parts of your recordkeeping system. We are interested in how you use it and why, not in any sensitive scientific information it contains. As you show us things, can we take pictures to help us remember? If there's something sensitive you'd rather we not take a picture of, just let us know when it comes up.

*Moderator: note a few features the participant mentions which are not already included in the Platform-Agnostic Feature Rank. These write-in cards can be used during said activity.*

## --- Process Deep Dive: Explore [5 Min, Total 30 Min]

First, we want to explore the beginning of the process, before you're ready to make a plan, when you're trying to figure out what you want to explore next…

8. What kinds of things do you generally do at this stage?
    - How do you build your research question or hypothesis?
9. Who are you working closely with at this stage? How do you collaborate?
    - What are the challenges with collaboration?
10. At this stage, what do you document? What can you look back at in later stages of the experiment?
    - What do other people see?
    - Can you tell us about a time when you needed to look back at something, but it wasn't written down?
11. How do you leverage previous work done by your own group or others?
    - *Probe other options if picture seems incomplete: replication? inspiration?*
    - What are you using? *(data, methods, just conclusions?)*
    - How do you build trust in that previous work?
    - What are the challenges in leveraging previous work?
    - What communication do you need or want to have with those involved in the previous work? *(data access? clarification? feedback?)*
    - How do you keep track of references or citations to previous work?

## --- Process Deep Dive: Plan [10 Min, Total 40 Min]

Okay, now you're ready to make a plan for your experiment…

12. What kinds of things do you generally do in this phase?
13. Who are you working closely with at this stage? How do you collaborate?
    - What are the challenges with collaboration?

- How are feedback and revisions handled?
14. Can you show us the plan from your most recent TRI experiment? *(ask this here, as later questions may want it as a visual aid)*
    - **[PHOTO]** *of the shown materials*
15. What makes a plan good? What purposes is it trying to fulfill?
16. What do you incorporate from previous plans or documents?
    - For things that are copied in, what kinds of modification is required?
    - What is referenced rather than directly included?
    - How are these copied or referenced things stored and managed?
17. What are some things you include in most or all plans?
18. It is our understanding that a workflow might be simple and linear, or complex with branches, repetition, conditionals… How do you manage this?
    - How do you represent and document this?

### --- Process Deep Dive: Execute [30 Min, Total 70 Min]

Okay, you have a plan. It's time to go into the lab and conduct the experiment…

19. Who are you working closely with at this stage? How do you collaborate?
    - How do you coordinate who is responsible for what?
    - For what do you and your colleagues ask each other for help while conducting the experiment?
20. Can you show us your notes from conducting your most recent TRI experiment? *(ask this here, as later questions may want it as a visual aid)*
    - **[PHOTO]** *of the shown materials*
    - *Observe:*
        - *If physical lab notebook, is there anything pasted into the physical lab notebook? If so, what?*
        - *Is there an organizational structure that allows searching/scanning?*
        - *Are there any free-hand drawings? Or anything else hard to captured electronically?*
        - *Are there any signatures or dates to establish the validity of the record?*
        - *Does it look like there is structure or placeholder space planned out ahead of time?*
    - Can you walk me through the important features of your notetaking system?
21. How did you come to this system for recording your notes?
    - What recommendations or requirements influenced that system?
    - In what ways has that system changed over time?
    - *(if using multiple formats)* Why do you prefer different formats for different information?
22. What do you do with these notes? What purposes do they fulfill?
    - Any editing?
    - Any sharing with others?
23. *(If relevant)* How do you combine your work with that of your colleagues?
24. How do you associate your notes with the plan steps?
25. Can you tell me about a situation where the plan document was missing something you needed to know?
    - How did you find what you need to know?
    - Did the information you find then get added to the plan?
    - How often does something like this happen?
    - How much of a challenge was this when you first joined the lab?
26. How do you account for when things don't go according to plan?
    - How often does this happen?
    - How do you document this?
    - How do you inform others?

27. Can you tell me about situations where you might do something in the lab without a formal plan? (*e.g., quick exploration*)
    - How do you document this?
28. Do you interpret any data while you're executing your plan, or only at the end once you've done everything? Why?
29. If a step in the process fails, what happens?
    - What, if anything, is documented about the failure?
    - Does the plan explicitly state how to determine whether a step failed and, if so, what to do? If not, by what means do you know what to do?
    - Did this happen during this experiment? How can I tell from the notes?
30. Can you talk me through how you manage samples and track their identity in your notes?
    - How do you name samples?
    - How do you and your group maintain information about samples over time?
    - How can you find out what happened to this sample before you received it?
    - How can you find other information collected about this sample?
    - What happens if you split or combine a sample?
    - What happens if you apply a processing step to a sample?
31. Can you show me some characterization data in your notes?
    - How do you organize data and associate it with your notes, both raw and processed?
    - Are there times where you need to look at the data, but don't need record or save it?
32. Can you show me some qualitative information you recorded in your notes?
    - How do you know when to record qualitative observations, such as this one?
    - How standardized is the terminology?
33. Can you tell me about a time where that was data or metadata you didn't document, but ultimately turned out to be important?
    - Is there any metadata you consistently capture? *(equipment identifiers, time, or environmental factors?)*

**[LOCATION CHANGE – Lab]** We'd like to have this part of the conversation in the lab, to better understand your context. Can you show us the way?

34. I want to better understand the physical context in which you are taking notes. *[Point to something in the notes]* Can you pose where and how you wrote this?
    - **[PHOTO]** *Take photo of them posing, understand spatial context for notetaking*
    - During this time, how were you referencing the plan?
35. What information was added to your notes…
    - …before you stepped foot in the lab?
    - …after you left the lab?

**[LOCATION CHANGE – Meeting Room]** Okay, we've conducted the experiment. We'll continue our conversation back in the meeting room.

### --- Process Deep Dive: Analyze [5 Min, Total 75 Min]

Okay, we've conducted the experiment. So, now you need to make sense of the information you've collected…

36. What kinds of things do you generally do at this stage?
37. Who are you working closely with at this stage? How do you collaborate?
    - When did you last analyze something collaboratively? What are the challenges?
    - What are the challenges with analyzing data collected by someone else?

- How do you get share and access data and notes?
- What interactions do you have with your advisor or PI?
38. What analysis tools do you use? Why those?
    - How have these tools or processes changed over your time in the lab?
    - Which tools do you use that are custom-made by your team? Why?
    - Rapid-fire, no-follow-up, questions:
        - Which programming languages do you use, if any? *(R, Python)*
        - Which databases do you use, if any? *(SQL)*
        - Which statistical analysis tools do you use, if any? *(Excel, SPSS)*
39. At this stage, what do you document?
    - What do other people see?
    - In what form?
40. Under what circumstances, if any, would you cut analysis short or skip it altogether?

### --- Process Deep Dive: Report [5 Min, Total 80 Min]

Okay, we've finished making sense of the information you've collected, we know what it means…

41. What kinds of things do you generally do at this stage? *(presentations, reports, publishing, intellectual property?)*
    - What challenges exist with remembering or finding what's needed to write the report?
42. Who are you working closely with at this stage? How do you collaborate?
    - When did you last write collaboratively? How? *(e.g., Divide and conquer? Emailing drafts? Collaborative platform?)*
43. What do you do differently if the results don't support your hypothesis, your goal for the experiment?
44. What do you do when the results conflict with prior work by your own group or others?
    - How does this conflict get added to the scientific record?
    - How do you discuss the discrepancies with those who conducted the prior work?
45. Who are the results communicated to, and why?
    - What is the recipient looking for?
    - What's your motivation for communicating these results?
    - In what form?

# Sharing Information [10 Min, Total 90 Min]
*Goal: Understand behaviors and attitudes on sharing information as both giver and receiver.*

46. **ACTIVITY #2 – Information Access Map**
    ***Goal:*** *Understand gaps between information access, perceptions of ownership, and actual ownership.*

    Over the full lifecycle of an experiment, there's a lot of information and knowledge you will produce and may make accessible to others. We've listed some of that information and knowledge here.

    *[Gesture to info/knowledge cards, such as: exploratory research, models and simulations, research plan, workflow, raw data, raw observations, analysis, conclusions, research paper]*

    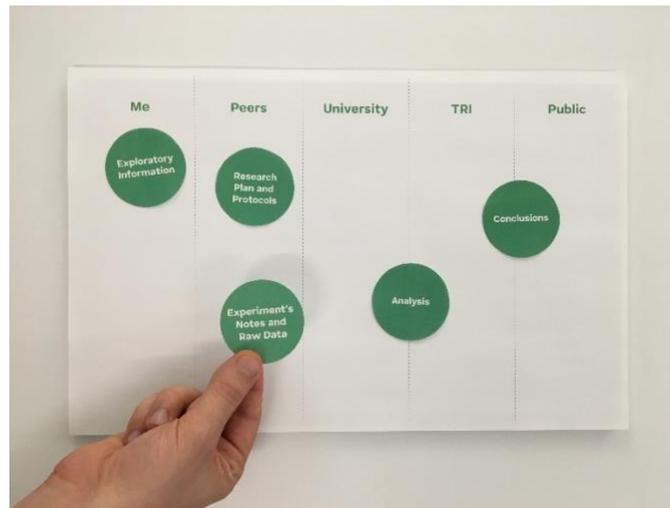

    *Sample of Information Access activity*

    Please place on the placemat who should have access to these pieces of information and knowledge and talk us through why.

    *For each item shared outside self: How? When?*

    **[PHOTO]** *of the results*

    When you are referencing someone else's work, which of these do you typically have access to? Do you wish you had more?

# Recordkeeping Preferences [15 Min, Total 105 Min]

*Goal: Understand what aspects of a recordkeeping system are most important to the participant, including perspectives on ELNs.*

47. We just spent a lot of time talking about your recordkeeping system throughout the lifecycle of an experiment…
    - When you first started in the lab, how did you decide how to keep records, what tools to use?
        - *Understand who influences these practices… e.g., self, peers, PI*
    - After you got your initial system in place, what changes did you make over time? Why?
        - What was the origin of the change? *(self, peers, PI?)*
        - What challenges did you face in making these changes?
    - What did you try out, but decide not to continue doing? Why?
    - What practices, if any, have you attempted to spread to others in your group?
        - How did that go?

48. **ACTIVITY #3 – Platform-Agnostic Feature Rank**

    Let's take a step back. We want to understand how best to improve your research experience.

    Take a minute to look at these aspects of your experience, then talk us through which 3 are most important to you and why.

    Of your most important features, what challenges do you have with them today?

    Which 2-3 are least important, and why?

    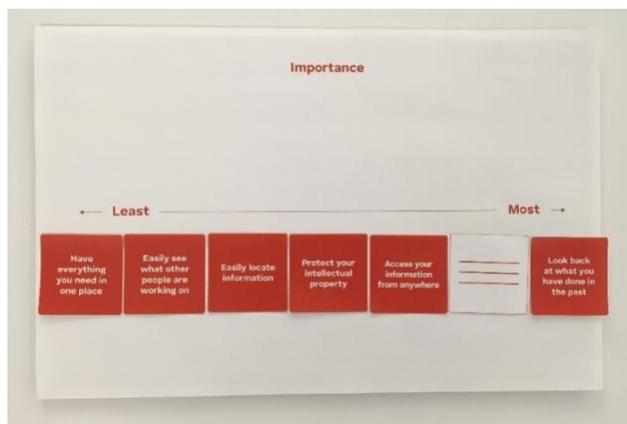
    *Outdated sample of Platform-Agnostic Feature Rank activity that conveys the point.*

    **[PHOTO]** *of responses*

49. What does an "Electronic Lab Notebook", or ELN, mean to you?
    - What is your perspective on ELNs?
    - *(If they have looking into one):* Why did you look into ELNs? What did you hope they would offer?
50. What is your experience with using an ELN?
    - If there's something you've used in the past but are using no longer, why?
    - *In reference to ranking from previous activity:*
        - Is there anything important from the last activity you think an ELN would help or hurt with?

# Current Experience Hypothesis [10 Min, Total 115 Min]

We put together a quick mock-up that we'd like your perspective on. We haven't made any decisions yet about what the right solution should be. This is just a conversation starter so we can better understand what would be useful for researchers like you. Your honest perspective is what will allow us to make the right choices, so please be candid.

- Project screen
    - Tell me about whether this feels reflective of how you think about the organization of your work.
    - If there was a way to search through your past work, what kinds of things would you search for?
    - How does this compare to how you think about organizing or viewing your work? What do you expect to see first?
    - What do you expect "View Workflows" to show?
- Workflow list
    - For the top workflow, what are your thoughts on having a list of tasks like this?
    - What do you expect to be able to do when you press "Record"?
    - When do you expect you would be using "Record"?
    - What do you think of this display of attempts?
    - What do you imagine this being useful for?
- Workflow details
    - When in your process do you think you would input a workflow like this?
    - What do you think about the information at the top about the workflow?
        - What might you use "Assign Experiment" for?
        - What might you use "tags" for?
    - How does this compare to how you think about what a "Step" is?
    - Talk me through whether this reflects how and when repetition occurs in your workflow.
    - [something about repetition]
    - When would you expect to enter your steps? *(before doing them? after?)*
- General
    - What do you not like about this? What concerns do you have?
    - What are you hoping something like this would help you with?
    - In what contexts would you expect to use this, and on what kind of device?

# Wrap-Up [5 Min, Total 120 Min]
*Goal: Address any lingering questions, conclude the interview, and thank the participant.*

Thank you for walking us through all the details of your experience, it's been very helpful. Before we wrap up today's interview, I have just a few last questions.

51. Is there anything else you think we should know?
52. Would you be interested in speaking with us again once we have some concepts to get feedback on?
53. **[PHOTO]** Can we take your picture?

Thank you so much for your time and honesty today, it was a pleasure speaking with you. Please be in touch if you have any questions at all. Take care.

**A user-centered approach to designing an experimental laboratory data platform**

Ha-Kyung Kwon[a]*, Chirranjeevi Balaji Gopal[a]*, Jared Kirschner[b], Santiago Caicedo[b], and Brian D. Storey[a]

[a] Toyota Research Institute, Los Altos, CA 94022, [b]Continuum-EPAM, Boston, MA 02210

*These authors contributed equally

Supplementary Materials - Participant Interview Debriefs

# Participant 2

Participant 2 is *"a synthetic chemist [who] designs the [materials] and [conducts the] synthesis."* Their goal is to create compounds that others can use. Participant 2 works with others to improve their understanding of their products. They now have a standard library of # compounds.

Participant 2 creates samples with multiple variations to figure out optimal state and what would yield better results.

Participant 2 does a series of experiments to verify the compound, *"a confirmation that you got [the] product [you intended to make]."* Typically, a project has multiple experiments and *"[multiple] compounds that are seeking to solve one common goal."* Participant 2 currently has # different projects.

Participant 2 *"makes the samples, but the measurement are done by other people in the lab."* Besides creating batches of their products, Participant 2 indicates ideal use conditions, what to measure, and a schematic drawing of of the sample – although *"it can be tedious to give this to others."*

Participant 2 receives processed data from their peers when they are unfamiliar with an instrument, an analysis technique, and with statistical information to predict optimal parameters for their compounds. Participant 2 exchanges data via email or dropbox. Participant 2 *"would take [others'] data product and take it from there for analysis."*

## Overview

### Raw data is only a stepping stone

Participant 2 uses processed and analyzed data to learn the properties, parameters, and behaviors of their compounds.

Participant 2 does not think other researchers on their team would need raw data to verify the nature of their products. Thus, verification is implied and trusted. *"Usually they don't care because this [raw data] is very detailed."*

NMR raw data is not saved, and Participant 2 mostly uses plotted graphs. Participant 2 later uses them to create the manuscript, but *"really [doesn't] look back at it [raw data]."*

### My plan can be unstructured and is based on a diagram

Before going to the lab, Participant 2 diagrams a targeted molecule structure. From the diagram, they get a big and small picture of the task at hand. Participant 2 explores different ways to achieve the desired compound by making slight modifications to the sample or conditions.

Depending on the familiarity with the compound, Participant 2 either copy-paste-edits literature when using a new material/structure or loosely defines plans when recreating a previously known product. *"Everything is here [points to reaction diagram]."*

Participant 2 does not plan ahead too much. They *"don't think is really critical to be included in the workflow, because it requires a lot of exploration... oftentimes it doesn't work.*

### To update my compounds, I need data I can understand

When others use and measure their products, they typically send Participant 2 *processed data* because it is more valuable and comprehensible to them. Participant 2 lets others characterize their compounds.

Many people work with their samples, so Participant 2 coordinates and aggregates information from several sources. The process can take time and require further analysis that Participant 2 carries on their own.

## EXPLORE

It is a communal effort to decide where to focus their research, along with literature and self-interests.

Intuition-based, and experience, dictates what [materials] to design. Participant 2 wants A.I. to help with predictions, *"That's the dream."*

## PLAN

Participant 2 *"[plans] things first. [They] need to design the [material] and how to synthesize."* This is an upstream activity that happens before exploring, and Participant 2 *"[doesn't] think is critical to be included in the protocol because usually things don't work."*

*"The plan is a synthetic scheme."* Participant 2 uses their chemical knowledge and literature to inform synthesis steps. Participant 2 would change the plan when experiments are not bringing any results or *"If it is too tedious, [they] give up."* Participant 2 creates the plan in chemdraw and shares it with others in digital format. These diagrams are the master plan, but they only print what they need to do for the day. Participant 2 refers to these plans until they have done it a few times.

## ANALYZE

Participant 2 does not need all the data points from others to complete their analysis, just the most important numbers.

Participant 2 usually verifies their samples and thinks others do not need these details. Sometimes, the process requires parallel analysis involving different researchers. *"Usually they don't care, they don't want to waste their time. It is very detailed [data]"*

NMR data confirms the success of their compounds, but they primarily work with *processed data*. *"Once I generated the processed data, I don't need the unprocessed data."*

## NOTE-TAKING

Participant 2 transcribes the plan for the day to create a compound from literature; It begins on the computer and moves into their notebook.

The product recipe sometimes does not get shared with others, but it always needs to be documented in the lab notebook. The recipe is very elaborated in the manuscript, but Participant 2 does not document basic assumptions on their notebook. *"If it is the first time doing it, there would be more text,"* and they typically pay more attention to what is changeable. Participant 2 shares synthesis recipes when they are not available and cannot assist, and the manuscript contains a neatly organized list of steps. They do not record everything, but *"your every day is in the lab notebook."*

A checkmark means that Participant 2 has done it or set up the experiment, and an "x" for when something fails.

## SAMPLE MANAGEMENT

Participant 2 organizes their sample-based on compounds and projects.

Their samples can be similar in structure, but Participant 2 tests them with slightly different parameters. Participant 2 names their sample with the date and page number of their notebook. They translates this information, along with relevant conditions, into file names.

When sharing their samples, they includes their initials, date, component information, and temperature range.

## DATA MANAGEMENT

Participant 2 learned to name their files from others at the lab, but *"sometimes it is difficult to understand others."*

There is no systematic way to organize shared data *"we do what's easiest for us."* However, Participant 2 doesn't think is a big issue.

*"Sometimes it is easier to go directly to the origin file"* instead of looking at a presentation *"origin files are only for me."*

## KNOWLEDGE TRANSFER

When working with others, sharing up-to-date data is challenging; it can exist in a shared folder, or be lost on email threads. Sometimes others update the graphs they are using, but the changes do not reflect on Participant 2's files. *"When working with multiple excel files, it is difficult to be sync on the same page. It could be better organized."*

Ideally, data would be a shared master excel sheet. Sometimes Participant 2 keeps two data sets. *"There is no organized structure that everyone uses."* Reporting happens through individual conversations.

Participant 2 does not save a *bad experiment* when it is due human mistake *"It could be that it's not made properly."*

## USE OF E.L.N.

*"I think it is a shared and searchable record that allows you to reorganize them in a way you want."* Participant 2 has not used an electronic lab notebook because *"I'm very familiar with the notebook."*

## CONSIDERATIONS

1. **See what team is working on:** Participant 2 does not need to see what others are working on.

2. **See what others have done in the past:** Participant 2 would prefer to talk to a person instead of looking through their data.

3. **Everything in one place** If everything in the team is in one place, it would be easier to find. *"I would not need to take each person's data and combine".* It's also good if different people want to see the same information.

4. **Understanding where I am:** I think I have a better understanding of where I'm in the workflow.

5. **Update workflow:** Not important because material synthesis is not phase-by-phase process.

## DATA OWNERSHIP & ACCESS

*Notes and Raw Data:* for now only me, but when published, it should be share with the public – *"especially if is profitable".* Notebooks are useful when patenting.

*Protocol and plan:* *"Useful to the person that worked on it."*

*Exploratory knowledge:* It would be more informative to know what's successful, or if nothing makes it successful.

*"Conclusions should be share with the rest."* Their main goal is to learn why, *"a fundamental understanding what generates good and bad properties in a compound. It is not about fully optimizing something."*

# Participant 3

Participant 3 works on characterization of samples synthesized by another group. Participant 3 tries to understand the connection between structures and properties of materials and their performance, such as stability or activity.

**ON FOLLOW-UP** *"I'd be happy to see [your concepts] and provide feedback, because we're going to be using this, so it would be better if we are making sure it is good."*

## Overview

### As audience widens, story becomes more self-contained

Participant 3 has a living PowerPoint document of their results. They share fresh content with their subgroup, who are aware of the project and general status. *"It's for analysis—as long as it's readable and helps discussion [it's good enough]."* With their closest advisor, Participant 3 may just show an Origin file and not even make a slide. Group meetings require more formal presentations which are concise, polished, self-contained stories.

*"[Group presentations are] more like a movie, [subgroup meetings are] more like a TV series."*

### Plans refined as they migrate; deleted when done

Participant 3 has a cloud-based note which they keeps up-to-date with their latest long-term plans and short-term to-do list. When something on the said note is completed, it is removed—there's no reason to refer to a completed plan. The note has only refined content. The raw content is usually in their small notebook which Participant 3 brings to meetings. There's no time to write refined content during these meetings. Instead, Participant 3 scribbles just enough for them to remember what happened as they refine over the next 1-2 days into said note.

### It's all about plotted data

Origin files are often the first form in which data is viewed by them and shared with others. Feedback they receive on plots is used to refine those plots. Sometimes, Participant 3 needs to re-analyze or re-plot data from a published figure. It can be hard to find this data. *"I want the Origin files."* They don't need, or want, anything more raw than that.

## EXPLORE

Generally, *"I want to know about the stability of materials, [and] want to correlate properties with materials."*

## PLAN

*"If everything's planned well, you don't have to keep thinking about whether I'm planning correctly."* Planning is *"intellectually demanding"* though not time-consuming, which is the opposite of execution. Metrics on which to evaluate materials are explored in literature and discussed with others.

They take quick, very brief notes in meetings in a small notebook. *"For this project, I'm going to do A, B, C, D."* They need to consider these notes afterwards to form a solid plan. *"I maintain my latest refined short- and long-term plans in a cloud-based note as a prioritized to-do list"*. Tasks may be scheduled for specific dates and times. Some things, like buying a sample or precursor, may take months. *"After I'm done with [a plan/step]… I just delete it."*

When ready to begin an experiment, *"I want to get everything prepared before I do some actions, so I don't have to keep going back and forth."*

## ANALYZE

*"If I know how to process the data, but have no idea how to interpret… [I'll] just show the raw figure"* to my mentor or lab members to discuss.

*"There's always new stuff coming out of your research. You want to understand it, rationalizing it takes a lot of time."*

## NOTE-TAKING

*"I always keep my lab notebook in the office or in the lab."* In all other contexts, Participant 3 uses a small, separate notebook, including experiments outside my lab, meetings, and conferences. Though *"I'd prefer to write things down electronically [because they] are kept better,"* *"I can't take [my] laptop into lab."*

*"There are always numbers or words I want to look at about my experiment that are very important that I need to keep checking… That's the point of the lab notebook."* *"I treat my notebook as more like a temporary… where things are, then eventually they will go to my computer."* While in the office, they may write a to-do list or calculations in their lab notebook. After a week of experiments, they'll take the lab notebook to the office and put the numbers in an Excel file.

## SAMPLE MANAGEMENT

Participant 3 will synthesize samples, then give others sample vials labeled with names and dates. They have an Excel spreadsheet tracking sample status and findings using the provider's sample name (e.g., for XRD: sample location, test status, qualitative sample purity).

## DATA MANAGEMENT

Raw data is stored in *ongoing_projects/experimental data/[instrument]/[date of experiment + sample]*. Within the [instrument] folder, they may have an Origin file plotting the associated raw data which they can open in discussions with others.

They also have folders for manuscripts and presentations.

## KNOWLEDGE TRANSFER

Participant 3 looks back at old presentations to: (1) understand what's new to share when making a new presentation, (2) when planning to see everything. "*I need in an organized way, including experimental conditions, parameters, data, and analysis. Notebook is more random, scattered. We want to wrap up in a logical way. That's why I make slides."* Anything that is important to analyzing new work will be added to a really long, living PowerPoint document, which is a reference for them or for discussions with those really close to their work. Presentations to larger groups further from their work are much more concise and formal.

There are two reasons to report: (1) *"when you're stuck, that's when we report, interact, discuss"*, and (2) *"on a time basis, particularly with PIs… after you collect a certain amount of data."*

## USE OF E.L.N.

Beyond the use of a cloud-based plan/to-do list, not specifically discussed.

## CONSIDERATIONS

1. **Quickly locate info:** when Participant 3 takes notes, they want to combine with related info from the past. But if they can't find it while notetaking, they'll note down what to look for. For data files, "*I think my current system on the computer is not bad, but could be improved."* For example, an analysis might combine data from two instruments, so they don't know in which instrument folder the analysis will be.

2. **See what I've done:** *"Things are very non-linear in my notebooks"* because they are working on multiple instruments in the same day or at the same time. Participant 3 may miss submitting a few things (in the prototype) unless there's an easy way to see what they've already submitted.

3. **Don't ask for too much**: *"there's a balance between the complexity and [time consumption]… it has to be concise but capture everything".*

## DATA OWNERSHIP & ACCESS

Participant 3 has *"mixed feelings"* about sharing notes, plans, and raw data. *"If people in the group keep looking at my plan, that's a bit creepy"* if it's still *"under development".* Once the plan is more concrete, it's fine to share. Notes and raw data *"should be shared if others want it, but… personal and work stuff is mixed, and it will make people wonder why they are looking at my notes."*

*"I've had to search through a previous post doc's notebook for synthesis conditions"*. It took a while to find because Participant 3 had no idea when [the postdoc] did the synthesis. They've also had to look for characterization data on another's hard drive, which *"may be harder, it's like a maze."*

# Participant 4

Participant 4 performs scientific research on topics related to the lab's goals: developing renewable energy technologies. *"I can do anything even vaguely related. I conduct experiments to determine the activity and stability of [materials] that I synthesize, and develop and improve synthesis methods"*.

"*We, the students, control what happens in the lab. We aren't told what to do, we do it ourselves*".

## Overview

### Recording is only for the official record

Not all information is meant for the official record. Exploratory information is excluded because, if it works, it will need to be repeated anyways. Setup debugging is excluded because it doesn't affect the science, even though having this information from others would help Participant 4 debug their own work.

*"Knowing that something [didn't work in a particular way] is only useful to know that the problem existed … but I won't use that data for anything… I wouldn't negatively be affected if I removed those pages from my lab notebook."*

### I'm in control, but want input from others

Autonomy is important to Participant 4.

The PI doesn't tell Participant 4 what to research, but does help them expand their thinking about research topics and questions. No one tells them how they should keep records, but provides them examples and conventions to consider and modify at will. Participant 4 understands that the problems they're facing now may have been encountered by others in the group (past and present). Participant 4 seeks to learn from their past experience and apply to their current situation.

### Growth is driven by in-person interactions

One-on-one conversations and group presentations provide feedback on progress, exposure to new ideas, and exchange of tips, tricks, and best practices.

### The plan is in my head

It's easy for Participant 4 to know what the high-level plan is and to track their place in it during day-to-day lab work.

*"I need to get data for all compositions in # intervals…. If I can't test one because of a mechanical issue, then I need to re-do it."*

## EXPLORE

Participant 4 will talk to experts to help them quickly find the right materials from the vast trove of literature. They dig into these initial articles and their references. Then Participant 4 speaks with senior members to make sure they are answering something new. They may repeat reference work to make sure their setup is good, asking experts for help if it's not working.

## PLAN

Synthesis can be a challenge. There's a starting point (protocol), but need to debug and optimize. It's not common to develop new experimental techniques in this particular field of research.

"*I plan to make multiple substrates in case something goes wrong or I think of more experiments to do.*"

The long-term plan is clear: test a range of compositions. The short-term plan might be written in their meeting notebook, or a form they can take to the lab like loose paper or sticky notes.

## NOTE-TAKING

The computer is the primary record. Anything important in the lab notebook is moved onto the computer, such as a formatted Excel spreadsheet or the filename of data from a lab instrument.

"*Our lab notebooks are kept pretty simple. It's mostly just recordkeeping of the experiments that you've done, with what sample, on what day, in what conditions. But usually all of that is kept also in the filenames of the data that we collect.*"

Settings (e.g., scan rate) are written down, but because they are standard across many experiments, they are only written the first time or when something changes. Qualitative observations are sometimes written, but often not worth moving to the computer.

There is no standard way of note-taking or record-keeping. Participant 4 created their own system informed by what others do. For example, their formatted Excel spreadsheet is a modified version of a colleague's, so it's easier to programmatically access their data.

### SAMPLE MANAGEMENT

Participant 4 has a container box with labeled slots to keep track of substrates. "*In my lab notebook, I write the container ID, the material, and the measurements I'm going to do.*"

### DATA MANAGEMENT

All important information is on the computer, and the filename is a crucial record. "*[It's] the most important [identifier]… we keep all the conditions, all the materials, everything you need in that filename.*"

Characterization is performed in another lab. They transfer data from proprietary instrument forms to Excel, then sends to himself via Google Drive for analysis.

## USE OF E.L.N.

I've never used a lab notebook software.

"*[An ELN should be] like a library, like a folder on your computer, and everything inside that folder is the official record of your experiments and data.*"

"*Right now, I have formatted Excel spreadsheets for data, PowerPoint presentations with feedback kept in slide notes, and Word documents for starting manuscripts. This is all saved on my computer synced with Google Drive.*"

## CONSIDERATIONS

**Learn quickly from the past:** navigating through literature to make sure you're doing something new is time-consuming. Literature also doesn't have everything they need in the paper or supplementary information. Participant 4 first talks to people in the lab and wishes they could reach out to alumni about their research for troubleshooting tips.

## ANALYZE

"*I talk to the PI, post docs, and other students about what I think the data means. They will push your thinking, help you ask new and better questions that inform your planning*". This can be in one-on-one conversations or via Q&A in group meetings.

Data must be processed before plotting, such as passing through a theoretical equation to get a meaningful output. This is done in Excel or MATLAB.

## KNOWLEDGE TRANSFER

Knowledge transfer primarily occurs through in-person interactions, either in one-on-one or group/sub-group meetings. Group meetings include presentations with Q&A, which are opportunities to share your data and get feedback. The presentation slides are mostly pictures of processed data plots with interpretation given verbally.

## DATA OWNERSHIP & ACCESS

Participant 4 thinks scientific papers should be open to the public because anyone interested in learning more about a specific topic should be able to do it for free. Even TRI funded research is published. This can include everything, including: notes and raw data; exploratory information, which Participant 4 thinks of as existing published literature.

However, this information shouldn't be shared beyond TRI before publication as there's competition to discover something first.

# Participant 6

*"How I mentor people depends on where they are in their career."* For several undergrads, Participant 6 selects and oversees their tasks, trains them on a technique so they can independently measure and analyze data. Participant 6 reviews their work via Dropbox or in-person so incorrect information doesn't *"propagate through the rest of our work."* For PhD students, *"I'm an advisor with expertise… [they] come to me when there's a problem to be solved"* or to discuss ideas/direction. Participant 6 also performs characterizations that they are an expert in for others.

## Overview

### Our time is valuable, don't waste it

Participant 6 feels organizational responsibility in addition to responsibilities as a researcher—their purview is the group, not just themselves. Participant 6 recognizes that people's time is valuable and they are careful with how they use it. Participant 6 only shares what is most important to their audience. Participant 6 focuses on doing the things only they are expert in, leaving the rest to others. Participant 6 is concerned that making more information available would take valuable time without much benefit to the research effort.

*"If we can cut corners that don't sacrifice the quality of what we're doing, we'll do that."*

### Missing something? Just re-do it

*"I do my best to understand my data without my lab notebook … If I'm going back [to my notes], I didn't think it was important. And if I didn't think it was important, I may not have written it down."*

Anything *really* important is always documented. If a trivial detail is not documented but ultimately needed for publication, it's not that bad. This happens once every few years. *"I just spend an extra afternoon redoing something. It's not worth the time to document details which aren't important."*

### No value in sharing failure

Failure means there's no valid data. For results that aren't failures but are just poor (e.g., lacks desired property), the results are documented but only sometimes shared. *"Nobody wants to hear a presentation full of 'I tried all these things, they didn't work.' [Those are only] if you're really having a tough time. I only talk to [the PI] about what has been working."*

*"It's not necessarily that the failure is a true result… it's a null result… a lot of times there are things that people have failed on that I can repeat and succeed at [and vice versa]."*

### EXPLORE

Participant 6 keeps a cloud document with all their ideas: for TRI, other projects, job applications, 20-year career plan, and ideas for others in the group to try. Participant 6 looks for things that are interesting, worthwhile, and fit in their tenure. Participant 6 keeps current on literature to understand where they can do something new and interesting.

### PLAN

*"I want to do as few experiments as possible to communicate whatever message I want to communicate."*

Their cloud document mixes short and long-term plans, such as: six formulations they want to run next week, a technique they want to try, or a research area that might be promising.

In lab, typically Participant 6 is mixing things together. They will make an Excel sheet calculating what they need for 5 different concentrations, print it out, and take it to the lab. *"Sometimes, I don't even save the Excel file."*

### ANALYZE

Participant 6 is very familiar with their techniques, so they don't need to spend much time analyzing. They extract info from raw data and put into a format suitable for presentation. They analyze their own data, and sometimes check the analysis of their undergrads. *"If I'm learning a new technique, I might do that together with someone to learn and become independent."*

### NOTE-TAKING

*"I do my best to never need a lab notebook to understand what my data is… If I lost it, it'd be fine."* Any relevant info is recorded the filename or settings file. My notebook contains setup information which is written once with only changes noted thereafter.

I may have a to-do list on loose paper or sticky notes, or a printed Excel sheet that I bring to lab. I may transfer it to the lab notebook. Molecules are drawn in ChemDraw then translated to the lab notebook, as I won't have the computer in lab.

*"The only time I've ever needed something not recorded was when I had to take over something from someone who had left."* Their procedure wasn't well recorded. We spent a year re-optimizing it. We eventually found it written on a Petri dish in a cabinet somewhere which had the numbers we recognized on it: 40 and 120. We tried the ratio, and it worked. *"Generally, when people are around, they are a better source of detail. Talking to people is more efficient than trying to find these numbers myself."*

### SAMPLE MANAGEMENT

Not important because *"a lot of what I make gets used up pretty quickly."* I might store a vial labeled with *"enough info for me to know about it, how much depends on how many others things I'm doing and how long it will be."* Otherwise, it's usually in my head.

In many cases, I've characterized others' samples *"where I didn't know what I was measuring, but it didn't matter. I labeled the file with [their] abbreviation. I didn't need to know. It was fine."*

### DATA MANAGEMENT

*"My data all goes into one place,"* a Dropbox folder. It's delineated by project, though that wouldn't be obvious to others. I split by instrument, then by material. One material can be associated with multiple projects.

*"I compile everything in Origin files… I have one for every project. Any useful or good data gets put in there."* It keeps everything in one place and plots in a nice way for reporting.

### KNOWLEDGE TRANSFER

*"100% of the time, [reporting] is a group effort."* We give presentations every few weeks to the university-TRI collaboration, monthly with collaborators, and weekly with the research group. In that order, we give increasing levels of specificity (data, methods, supporting info). I report via presentation internally everything that works, even if it performs poorly, but never include it in a paper. *"I go back to PowerPoints a lot… those are the most distilled, other than the papers, of what I do. If I want to remember what I was doing on this project, what the data looked like, I will go back to the PowerPoints first."* When writing a paper, if *"[another] group did NMR, I don't care about that info, I just know they have it. I ask them to put in [the details], they add it and send it back."*

### USE OF E.L.N.

Even if provided a lab computer, Participant 6 would still be concerned about bench space, portability, ability to write freeform, and worse organization despite searchability. *"I found a failed [sample] in a lab notebook from 4 years ago quickly [just now]. It's more tractable to look at someone's old paper notes."*

### CONSIDERATIONS

1. **Data backup is critical:** *"losing all my info terrifies me."* They have their data in DropBox and on 2 different hard drives.

2. **Need for multiple views:** It's easier to copy/paste instrument data if it's organized by instrument, but better to share with others by project.

3. **I don't care what others are working on:** *"I know I'll get an update from them [at meetings]... it's not worth my time."*

4. **… nor about what others have done:** *"I understand that the point of this [prototype] is something that lives beyond me… [but] I've never looked at data that an old lab member has taken, except [in rare instances], but that [data's] already on the instrument... I don't think I would ever go back to look at data on [the prototype]."*

### DATA OWNERSHIP & ACCESS

*"I've never looked at anyone else's lab notebook."* *"Nobody wants to look at this, it's nonsense for most people"* Whatever is useful ends up in data file or slides.

Regardless, I am willing to share raw notes and data as long as I know what it's being used for, but it would be a non-trivial amount of work to make it understandable. I have two filters for deciding to share: (1) *"Do I trust you?"*, and (2) *"Is it worth my time to share?"* I trust TRI more than other research sponsors, but I am not sure why this would be a valuable use of time—*"that's a conversation to have."*

# Participant 7

Participant 7's goal is to pass the synthesis technique they had learned from their mentor to others on the team. Participant 7's *"primary focus is learning and further developing synthesis methods and new materials"*.

Experiments take longer than they initially expected, since "*synthesis takes a week or more, and [...] characterization takes another week. Each batch is an intense process."*

In a typical week they typically do 2-3 synthesis in a day, twice a week, for a total of 6-8 samples per week.

Their mentor works with Participant 7 in the lab. Their mentor passes the knowledge of synthesis work and provides general training for lab work.

Participant 7 has learned a specific synthesis technique from their mentor and will be *"fully be taking over the mentor's work."* They prepare samples together, and *"together collect data that [Participant 7 and the mentor] both analyze on [their] own."*

## Overview

### I work with many samples at once, and things get hairy

Participant 7 produces multiple variations of a single sample. They want to learn how the samples perform in different conditions or to test what they sees in the literature.

Because a sample can have multiple iterations, Participant 7 wants to keep track of how samples evolve as others use these samples for new compounds. Synthesizing and characterizing a large number of samples forces Participant 7 to not record every detail. One thing is clear: when things do not work, Participant 7 stops recording data.

### The use of a lab document is encouraged, yet I don't use it

Their PI has promoted the use of a shared lab document, but most people do not use this document. Participant 7 has contributed little to it because *"the problem is that [they] would have to type everything in full"* and they question: *"If it is in the lab notebook, why would it have to be in the lab document?"*

Participant 7 thinks a shared lab document is good for results, but it cannot replace the notebook. It might help others, but it would not help Participant 7 since *"raw data is only useful to [themselves]"*

Even when tools are expected to be used by researchers, PIs are not verifying or checking their usage.

### I'm protective of what's novel

Because their lab is using specific materials and techniques, Participant 7 is wary of what is shared with others. They believe some synthesis information is proprietary to their work because *"methods have trade secrets."*

## EXPLORE

Participant 7 is currently looking at a wide range of articles because they are working with materials they have not had experience with. A challenge when using literature at the beginning of an experiment is to *"figure out what might be a challenge in the future."*

Participant 7 has their library of articles. One of the issues they have is to reproduce exactly what they find in literature. Participant 7 uses printed standard protocols that they keep in their notebook: *"I've annotated on my own."*

## PLAN

"[sometimes] *it is a 12 hour day, and that can be an intense day. I have to have everything prepared in advance."*

They draw characterization diagrams before going to the lab. Because synthesis of a material involves several tweaks, their plans change on-the-fly as the experiment evolves.

## ANALYZE

Participant 7 uses analysis tools to understand whether a material is behaving as expected. They go back to planning whenever they discover that an experiment is not going as expected.

Analysis happens on a computer. However, this analysis process can also start from annotated articles, or during the execution of an experiment.

To report their progress, Participant 7 uses graphs, but does not use *"all the data points from the origin file."* They report what is useful, *"not all the numbers that came before the conclusion."*

## NOTE-TAKING

Because Participant 7 *"[is] always in and outside the glovebox,"* taking notes during synthesis is challenging. Participant 7 creates handwritten tables with observed results because they *"cannot take [their] laptops to the lab."* They draw characterization curves before going to the lab *"because synthesis is similar, I write everything that is particular to that, and the rest is write down as is performed."* They annotate on their notebook all the steps to create a standardized material that they learned from their mentor.

Participant 7 usually does not take their lab notebook out of the lab but has a separate notepad to analyze results at home. Failures are not recorded: *"I don't usually write down here what went wrong, I generally know when things go wrong."* Additionally, they keep a ruler on their notebook to measure experiments, but it *"also works as a bookmark."*

They go back to their notes to see what happened in the previous experiment; They also review previous procedures and materials that they had developed. Typically they would use the notebook to enter what happens at the lab and share it via presentations.

## SAMPLE MANAGEMENT

Each material *"has children materials"* and a child can become a new material due to the synthesis process. When using the furnace, they only put/use similar materials. A typical sample has more than 5 children samples.

Participant 7 names their samples depending on the number of the product. Some samples are personal, and others are shared *"I'm in my synthesis ##."*

## DATA MANAGEMENT

They name their files with their name, synthesis number, and some conditions.

*"Everything is saved to my laptop"* They keep sample names in sequential order. Participant 7 saves files in a shared drive, and access them remotely from their personal computer. They would rather have data files on their computer before using analysis tools. *"Everything is stored somewhere and I'm able to retrieve everything I need"*

## KNOWLEDGE TRANSFER

Participant 7 learns from their mentor by doing procedures with the mentor at the lab. *"I am fairly independent, but I am still learning the details."*

In regular subgroup meeting, *"that is where conclusions get explicitly defined."* If they get lost on what they are doing *"I look back at the subgroup meetings, and learn "oh that is why."* They revise presentations to learn the status of a project; Participant 7 wants to make sure that they are on track to accomplish their goals, and to show others that they did their job..

Participant 7 reports when their experiments do not go as expected *"as part of the subgroup, and I usually think about what went wrong with an experiment, so I have a hypothesis ready."*

## USE OF E.L.N.

Participant 7's version of an ELN is a shared document. *"I think they are a good idea."* In practice, this resource is not valuable because its use is inconsistent: *"In theory we are supposed to use the shared document, in practice, we do not"* The lab document was developed a few years ago by, *"but is something that is never checked or verified."*

If information exists in the lab notebook, *"why would it be in the shared document?"* They would use the shared document for results, *"but if I had to use it for everything, it would be a time drained."*

## CONSIDERATIONS

1. **Keeping track where I am:** *"I'm aware of where I am."* Participant 7 likes the idea of adding information about the part of the process they are at.

2. **Looking back at others work:** *"it is not as important because what I'm doing is quite unique, because it has not been done by others."*

3. **Backup files:** Is not important because they already have dropbox, shared drive, and cloud storage.

## DATA OWNERSHIP & ACCESS

The shared lab document contains standard materials and its primary functions as a 1) repository of information, 2) is helpful for patent purposes and 3) is time-stamped. *"I do not use the document to its full capabilities"* because Participant 7 has too many samples to record, *"having to go and type everything is entirely is time-consuming."*

Details of Participant 7's experiments only belong to them, and they do not think that this information would be interesting to others. *"Research plan belongs with TRI, but on a day-to-day basis, it belongs to my subgroup."* Reports include *"What worked, what did not work," "You probably do not need to know every synthesis that did not work (…) I do not think TRI needs that."*

# Participant 8

As a postdoc/research staff, Participant 8 is known to be a second voice whenever the PI is not present: They send emails on the PI's behalf, and sometimes teaches the PI's classes. Participant 8 also contributes to writing proposals and updating the PI about researchers.

Participant 8 collaborates with multiple groups outside of the university. Their role is to explain to non-technical audiences a digestible version of the relevant research.

Participant 8 mentors 15-20 students at any given time and helps the students narrow their scope of research. Because _"[Participant 8] cannot tell a student what to do or not, sometimes the students spreads a little thin because they have many projects."_

## Overview

**I only share my progress when there's an interesting story**

Participant 8 processes their data and plots it into graphs. They place the graphs in draft presentations so that they can see progress in one place, and to craft the right story to tell others.

Experimental raw and analyzed data are mostly for Participant 8, who _"would never share [draft presentations or raw data]. This is just [Participant 8] working out what's going on and deciding whether to throw this data away"_

When there's no good story to tell, they re-think their next steps and craft a new plan to gather more data. _"This is a lot of data to show others [in the presentations], and [Participant 8] just cannot be bombarded with this much data in a presentation until [they] get the story straight."_

**ELN makes my life easier, but it won't replace a lab notebook**

Participant 8 doesn't bring their personal laptop to the lab because they are concerned with chemical safety.

An ideal ELN would combine planning and logging. It needs to be low cost or free, enables collaboration and usable outside the lab. Participant 8 sees an ELN as a way to record by-sample data and PowerPoint for by-project analysis. An ELN minimizes data entry for multiple samples processed by the same technique with few parameter changes.

Working in the cleanroom made Participant 8 use an ELN because the fiber in notebooks could interfere with some sensitive electronic instruments. This situation converted them to notetaking electronically.

**Lab notebooks are personal, and I don't need to see others**

Substantial effort is put into planning and conducting experiments. When experiments fail, lab notebooks become personal artifacts because they retain these emotional moments.

To make sense of a lab notebook, Participant 8 thinks lab notebooks can only be understood by its creator. Otherwise, a viewer would have to learn and decode the author's system. All in all, Participant 8 does not need to see their mentee's notebooks.

If there was a case where Participant 8 had to hand off their notebook, they feel comfortable sharing it but would like to know in advance so they can structure in a way that makes sense to others.

**MOTIVATIONS** _"The motivation has always been environmental impact. And as I'm getting older, I'm a little less sure that my day-to-day activities will impact anything, but I like teaching, and that's something I can contribute._

_"We are interested in fundamental of science. We want to understand how it all works and make some rational observations"_

## EXPLORE

Given their mentorial and managerial role, Participant 8 is mostly helping new researchers to focus on the right questions to explore. Participant 8 also directs the students to a starting point: literature or experts in the field.

Participant 8 thinks exploratory research can be pre-determined by funding: *"TRI tells [the group] what they are interested in, [and the group] explores and plans but within some limits… it is limiting, but that is how it is in academia."*

Participant 8's group works with other industry-known collaborators that are interested in similar areas of research, and researchers are working across different funding partners.

## PLAN

When crafting proposals, Participant 8 helps to establish the goals and lays out the foundational work for the research plan. The specificity of the plan depends on its duration, novelty, and complexity: "*sometimes [the researcher] has to plan down to the month and what [they] will achieve. And the experiments have to align exactly. That's one extreme*" and the other is a more flexible plan that changes based on new findings.

For experimental work, Participant 8 does not have a concrete project plan, just a daily plan or to-do list. Participant 8 starts electronically on an ELN, and moves it to paper but their *"favorite is post-it notes … because the lab is disgusting"* and post-it notes can be discarded.

Participant 8 bases their detailed plans on literature by "*[copying and pasting plans] directly from literature.*" Unstructured or loosely defined plans are based on intuition or previous experience. A good plan is "*a starting point, and some kind of [a] goal.*"

## NOTE-TAKING

Participant 8 lab notebook captures what happened at the lab, since they *"just go and measure in the lab."* Participant 8 prefers paper for its portability and flexibility to accommodate different type of notes, such as drawings and checklist on post-its. Paper is *"less trying than typing."* Most importantly, Participant 8 likes how easy is to take notes on-the-fly, since *"it is just faster and less bulky to have a notebook and [be able to] just scribble scribble scribble."* Participant 8 tries to transcribe all their lab notes into presentations and names of files *"to be reminded of the [experiment] conditions."* Participant 8 also annotates instructions for what to do in the lab (for using instruments) and for things they would like to remember.

The lab notebook is a personal object that can be "*a little embarrassing.*" Participant 8 would structure it differently if others were to see it because *"If it takes 5 days to make a sample and then in 10 seconds you see it didn't work…there's a lot of emotions in that I do not want to share that broadly"*. Participant 8's mentees have not shown them their notebooks. In order to make sense of others' notebooks, Participant 8 would have to learn their system *"because I cannot see the data (…) and that would probably be a minefield."*

## SAMPLE MANAGEMENT

Participant 8 keeps track of the name of the sample, in addition to the date and the conditions. Participant 8 names the samples in a chronological order. Participant 8 "*really [likes] long file names [that contain] all the conditions of the experiment [so that they] never need to worry that [they] are missing something."*

Participant 8 doesn't like to name samples by numbers because "*in a week, [it's difficult to remember] what they mean.*"

## DATA MANAGEMENT

Participant 8 can read other group members' file names because the file naming convention is common across the group. Participant 8 learned this from others in the group, and *"train[s] students and enforce[s] them to use this system."* Participant 8 *"explains why [they] did it and why [they] think it is good."* Participant 8 has never seen others' raw data.

When looking for presentations, Participant 8 has a hard time finding them because they have *"done so many presentations."*

## USE OF E.L.N.

Working at the cleanroom made Participant 8 use an ELN because the fiber in the paper could interfere with sensitive electronic instruments. Participant 8 is concerned about chemicals and electronic note taking.

Participant 8 has used multiple electronic lab notebook softwares. Participant 8 uses electronic lab notebook software that has a co-authoring feature Participant 8 needed to work with their colleagues. Participant 8 "*like[s] the search function to look for keywords.*"

## CONSIDERATIONS

1. **Know where you are:** *"I don't feel like I have a well defined workflow in the lab. It would be hard to define a workflow."*

2. **Know where you are :** Participant 8 does not know whether the knowledge of knowing where you are would be helpful. "*If you stop me at the lab, I would know where I am, and I might know the next step.*"

## KNOWLEDGE TRANSFER

Presentations are milestones for Participant 8 and their project because the slides reflect confidence in Participant 8's work and contain the most relevant data. These artifacts are more than timestamps; Participant 8 values how the slides highlight the progress of the project and how *"it forces [them] to be more organized and think about what [they] have done."*

Participant 8 looks back at presentations to connect what they have done with what they are doing, see what happened in a meeting, or simply re-utilize the information in the presentations. *"If [a researcher] already plotted all [their] data, it is easy to copy and paste it into formal presentations."*

Whether or not Participant 8 can interpret someone else's notes are not important, as *"presentations are [more easily understood equivalents] of lab notes"*

## DATA OWNERSHIP & ACCESS

Participant 8 prefers for others to see the slides after they are sure of the story. Participant 8 *"[would] never show [draft slides] to others, because this is [them] working out whether or not to throw this data away."*

Sharing data with a funding partner is not an issue, as long as they do not micro-manage Participant 8's research. *"Raw data and experimental notes are good to share with a collaborator who wants to [validate their] data, [but there] needs to be a balance in getting feedback and a fear of others accessing the data."* Sharing proposals or data with a competitor before publishing is a no-go, since *"[researchers] spend a lot of time and effort on this…that would be heartbreaking."* Once a manuscript is published, everything else can be made public.

# Participant 9

Participant 9's role is the chemistry—to figure out what would be the next best system to try to make. The materials Participant 9 synthesizes are characterized by others.

Participant 9 also helps with other projects as a resource—someone who can synthesize things when they come up as ideas.

## Overview

### Lab notebook page is the unit of organization

Participant 9 uses one notebook page per experiment. Everything about that experiment is named and organized by associated notebook # and page #, including data filenames and samples. They have even tried keeping PowerPoint slides of analyzed data where slide number maps to notebook page #, but that turned out to be impractical. An experiment is defined as isolating a synthesized material (or a material set varying on a parameter) and having measurements taken on it. Information in Participant 9's lab notebook has a longer useful life than others we spoke to, perhaps related to Participant 9 not actually using it in the lab.

### Context switching is a challenge

Participant 9 is working on multiple projects in parallel and may only have small pockets of time to work in lab between meetings. It can be difficult to quickly figure out where they left off with a project and set up the next experiment. Beyond memory, they rely on notebook pages and presentations in which projects are interspersed.

"*…it can be hard to think critically about something you're working on when you're trying to split your time. Sometimes, you'll get a result that is unexpected [that] requires you to think about all your previous knowledge…*"

### Raw data is not to be seen

Participant 9 creates and shares their interpretation of raw data. Only if something doesn't look right should anyone look back at raw data, including themselves. Access is still important to enable plotting others' data to your own preferences.

### Work is solo, with guidance from others

Advice, feedback, and expertise is sought from others, but day-to-day work is solo. Participant 9 synthesizes materials characterized by others with limited interaction: material given, measurements received.

*"You make this, I'll make this, we'll put it together for a final project."*

## EXPLORE

Once the group has an idea of what material to make, Participant 9 and their group mates dive into the literature to figure out how. This can be tedious, but *"3 hours at a computer will save 3 days in the lab"*. Participant 9 saves their references in a referencing software and sometimes in their lab notebook.

*"A lot of times, you just do test reactions. Maybe this is the first time with this specific compound. You're trying to figure out the small details of this reaction and optimize it."*

## PLAN

Everything Participant 9 does in the lab is predetermined. The lab notebook with the plan stays in the office. Whatever is needed as reference in the lab is printed or copied to sticky notes.

*"I set up all my reactions… write them all out, do everything I need on the computer… then print out a piece of paper with all the reagents I need to add."*

## ANALYZE

Participant 9 extracts instrument data to a CSV or text file, copies it to a flash drive or emails to themselves, then works up in some program. Ideally, Participant 9 confirms results using several instruments. Most data are presented as XY scatter plots with very little data manipulation required. The techniques were learned in school and are widely known.

Participant 9 may iterate 10 times through explore/plan, explore, and analyze before reporting anything.

## NOTE-TAKING

*"The lab notebook is supposed to be with you at your bench, as a reference… [but] I don't like to bring it into the lab and then back into the office."*

Instead, Participant 9 records data on printed paper or sticky notes in the lab, then takes them back to their office (immediately adjacent). Participant 9 copies these notes and other things into their notebook or Excel. If Participant 9 has a lot of data and uses Excel, Participant 9's notebook will reference this. Their notes help them remember reactions, setup, actual reagent amounts. Only notes for experiments that worked and will be published are moved to an electronic format. It would be good if Participant 9 recorded the reagent purity and vendor in the notebook as they will need it for the manuscript, but it's usually from a big batch so they "*can just go back and look if [they] need it*."

*"There's not usually someone saying you have to keep a good lab notebook, it's… a self-motivated thing."*

Participant 9 may have multiple projects and reactions going in parallel, so notebook pages can be interweaved. They *"don't have a system for linking projects… it's mostly by memory."* Participant 9 tried using project names, but the names need to be really long to be useful. For example, "[Material]-[description]-[purpose]" is initially descriptive, until they have a second project with the same "[Material]-[description]-[purpose]".

## SAMPLE MANAGEMENT

Each experiment is a synthesis, and each synthesis has its own notebook page. Samples are identified by the notebook page containing their synthesis recipe.

*"Right now, if I have a reaction that works, I remember because it's so exciting. I remember from grad school certain notebooks that were exciting."*

## DATA MANAGEMENT

*"We're all working on several different projects, and we don't really have a system right now that keeps track of … the last experiment we've run on a project… you just have to remember, or hopefully you have a research update."*

Participant 9 has a folder of these presentations by date. If they want to find information from the past, Participant 9 may have to open a lot of presentations to find it. They might look in their notebook for a related page with a date to help them find the data.

*"If I'm looking for data, I will look at my presentations."*

## KNOWLEDGE TRANSFER

Participant 9 shares results once they are confident in their conclusiveness. Participant 9 takes all the data for the conclusive reactions they have done each month and condense it into a PowerPoint presentation. If # attempts were made to create the same product, they would only show the information from the reaction that worked (or worked best).

*"[The lab notebook] is a very valuable thing when you're trying to make presentations."*

*"When you're presenting to your boss, you don't have the lab notebook… This is all the nice data to demonstrate that it worked."*

## USE OF E.L.N.

Participant 9 already uses their lab notebook exclusively outside the lab, where they have computer access. To Participant 9, an ELN is a lab notebook but on a computer. Participant 9 has tried a few ELNs, but they abandoned them because none had a way to draw chemical structures or places to keep data. While they are pretty sure these features exist, they have never tried ones that weren't free.

*"I've been dying to replace my lab notebook. I realize how inefficient it is. I don't get access to this after I leave an institution."*

## CONSIDERATIONS

1. **Quickly locate info:** *"If I could open up my notebook and have all the data there analyzed, maybe some conclusion sentences about this data, that would be extremely useful, versus finding in my lab notebook and presentations and making conclusions that way. That would save a lot of time."*

Participant 9 struggled to find or make a system like this for themselves. And, if they "*can get to [their] desk and know exactly where to pick up next, that can be useful.*"

## DATA OWNERSHIP & ACCESS

Participant 9 has not looked at someone else's notebook in a long time. *"It's kind of taboo to do that,"* although they did in undergrad. *"In some research environments, it's kind of competitive."* But if they need information from a groupmate who is unavailable, they will look at the group mate's lab notebook.

The lab notebook is personal and casual, with notes just for Participant 9. *"If it's going to be published or shared [with TRI], I'd want it to be more nice and neat".* It also has mistakes, but *"chemists get over that pretty quickly because it's time consuming. [Chemists] should be running reactions, not drawing stuff."*

# Participant 10

Participant 10 mostly works with two individuals. (1) A senior person who is particularly involved in project definition, analysis, and writing. When Participant 10 started, the senior person gave them a list of ideas to investigate, but now, Participant 10 comes up with ideas and the senior member vets them. (2) Participant 10 sometimes collaborates with another senior member on execution, particularly for complicated experiments. The PI is only involved in the report. Participant 10 meets in-person monthly with other groups that they are collaborating with.

## Overview

### Notes need to move with you, but the lab notebook can't

Participant 10 has two notebooks: (1) a large one for use in the lab; (2) a small one for everything else, including plans, ideas, and discussion notes. Outside the lab, Participant 10 may write in the small notebook, loose paper, or sticky notes, and then transfer it if needed to the big notebook for use while in lab. Standard protocols are printed and then stored in the lab notebook, taken out of the lab for reference and modification as needed.

### Presentations show you where you are in the process

The presentations Participant 10 gives at group meetings are what they will refer back to later to understand their progress. These presentations are primarily focused on showing result and not on intermediate steps, though they don't know what TRI actually expects. What matters is how exhaustive the data and understanding of the data is.

*"Because we make presentations pretty frequently, that's usually where I go back to… I had a presentation a month ago, this is the data that I had at the time."*

### No one but me understands my raw notes and data

Participant 10 isn't concerned about sharing raw notes and data, they just think it would be pointless as it wouldn't be understood. Participant 10 doesn't really share notes with others, though they might write things down to talk about with others in-person. Participant 10 has never looked at someone else's lab notebook.

*"If my [file] naming scheme isn't very understandable to someone I'm working with, it's probably not understandable to anyone else."*

## EXPLORE

Participant 10 reads research papers, talks to people in their group, and try to figure out what's been done and what's feasible to do. Participant 10 keeps a copy of downloaded papers on their laptop, and may have a printed out version with annotations.

## PLAN

Participant 10 may plan basic experiments to test quickly whether it's worth pursuing further. If Participant 10 references another paper, they normally try to replicate what the authors did to make sure that they have a solid foundation. The most intensive planning happens for large, collaborative experiments. Though the group plans weeks ahead of time, _"everything goes out the window when you start, because nothing ever works."_

Participant 10 generally knows the overall plan without writing anything down. Bigger picture ideas stick with Participant 10 better than the nitty-gritty. Participant 10 has two notebooks. (1) small and portable, used to: plan experiments, ideas they want to explore more, discussion topics and notes, weekly plan. (2) lab notebook that lives in the lab. Participant 10 typically cross-references the two by date.

## ANALYZE

Analysis is complicated. There's a lot of data. Participant 10 tries to figure out how it agrees with itself or doesn't. You can think you know what's going on based on all your experiments, but until you write it down, you don't know. You may find that you need more experiments. This is the most important step to Participant 10, because you don't know you've solved the problem until you do this. For straightforward things, they can decide on their own whether they've solved it, but for more complicated ones, they need to talk to other people.

## NOTE-TAKING

The lab notebook is for qualitative or observational data. Entries are organized by date, which links these notes to data files. Before Participant 10 starts working in the lab, the page may include what they want to test today and some calculations, such as how much material they need to weigh out.

Participant 10 synthesizes samples, gives them to someone else to characterize, receives the images back by email, saves them, and adds to a presentation. If the images show something wrong with the synthesis, they will change parameters and repeat.

Gaining maturity as a notetaker means _"remembering what you are not going to remember, and writing it down."_ For example, Participant 10 has notes from safety training that they wrote down when they started and still refer to.

## SAMPLE MANAGEMENT

Participant 10 has sample boxes with 2D grids of samples named according to their box number and grid location (e.g., B#R#C# is Box #, row #, column #). Participant 10 has an Excel spreadsheet with one tab per box, and grid cells within that tab map to box cells. Cells in the Excel file have text describing synthesis (composition, synthesis conditions, date) and color indicating measurements taken. Some measurements change the sample, so they do not use that sample again. Sometimes, Participant 10 gives a sample they made to someone else to characterize.

## DATA MANAGEMENT

Data files are stored on a network drive and Participant 10's computer. The file names contain a lot of metadata, such as a date to cross-reference with their lab notebook if needed.

Information is shared in-person, even if accessible electronically. _"It's complicated when one of the team is on vacation. Our naming schemes are a bit different. The information is all there, but it can be hard to figure out what it means,"_ or even where it is.

## USE OF E.L.N.

Participant 10 has used a commercial ELN in the past and liked it. It keeps things organized, in one place, in date-order. It would make sharing easy, but they never did that. That said, they have switched back to hand-written notes.

_"I remember better when I write down versus type."_

## CONSIDERATIONS

1. **Everything in one place:** _"I have some challenges with that. I try to keep track of everything, but cross-referencing can be difficult"_
2. **Flexibility:** _"Being locked into a particular workflow is bad. I want to be able to change my mind on how things are going without it being difficult." "I would like something flexible, not a bunch of fields you have to put in, but the things you want are there"._
3. **Tracking:** _"If it's where I am in the experiment, that's something I know, I don't need help… If it meant at a more project level, that could be useful."_

## KNOWLEDGE TRANSFER

Participant 10 gives presentations regularly. These slides aren't just for others; Participant 10 refers to them to understand their status and progress.

When Participant 10 collaborates with others outside their group, _"[the outside group is] interested in the results, the analysis. They don't care about the details of what I did… things I need to do right but only matter to us."_

In reports, there's a new push by funding agencies and journals to publish all raw data that went into figures.

## DATA OWNERSHIP & ACCESS

Participant 10 only shares their notebook in limited cases, generally with someone in their group. _"The team should [and TRI could] have access to experimental notes and raw data, though I don't think they'll need or want it."_

In general, Participant 10 believes that the general goal of academic research is to _"increase the body of knowledge, not just for [academia] but for everyone,"_ but they don't see how sharing the raw data would be useful. They need to share in presentations, papers, or patents to be recognized; it's not about owning the data.

# Participant 11

## Overview

### The E platforms has value as a manuscript builder

From Participant 11's perspective, the prototype is addressing two different issues 1) as an ELN to record data and 2) as database and file repository. They believe that the solution should assist them with the creation of manuscripts before publishing; The solution should not support o-going documentation of the project. *"A published paper is the same as this [pointing to prototype]"*

For Participant 11, not many solutions could replace a paper notebook.

### Loosely defined plans for an unforeseeable future

Participant 11 does not have certainty that their experiments will playout in the way they expect – in fact, *"most do not succeed."* Their steps are not linear, but Participant 11 has an idea of where their project is heading and what they need to do next.

Given the serendipitous nature of their work, Participant 11 does not do forward-planning, *"I am not going to plan when I do not know what I am going to end up with."* Their workflows are high level.

### A structured onboarding could set the right foundation

There is no official onboarding for new members of Participant 11's research group. The lab know-how comes from seeing others doing it, but Participant 11 would like more directed guidance from mentors and PI.

Participant 11 would follow standards for notetaking and file management – if they existed from the get-go.

## Researcher's Context

**WHO DO I WORK WITH?**

Participant 11 is part of multiple research groups, which means *"just more meetings, and trying to learn from both sides."* They work directly with a mentor characterizing materials, and receives guidance from other senior members. This is good, since *"[getting] thrown into a lot of things is good for learning from many different people."*

Joining a research group is a tricky part for grad students. They are inclined to learn what they are not familiar with, and wish that others made *"introductions to techniques and one-on-one time for guidance."*

**WHAT'S MY RELATIONSHIP TO TRI**

Reports happen every quarter, and Participant 11 provides an update of their work. Participant 11 starts with a few paragraphs in a word document. Then they send their part to a collaborator, who edits their part and sends it to a senior member who coordinates the writing of the report.

Participant 11 understands TRI's overarching goal, but feel that the expectations are not crystal clear. Participant 11 thinks that more direct interactions with TRI could be fruitful because *"all [their] projects are noble approaches, and [they] think it would be interesting for TRI to see."*

**WHAT'S MY WORKLOAD?**

Participant 11 *"has [their] own little projects."* Participant 11 spends approximately 70% in one group, and the remaining time with another group.

*"As I make new products and characterize them, I will be working more closely with both groups"*

## EXPLORE

Participant 11 prefers to ask their colleagues how to use instruments because it is _"better find someone else that can walk [them] through."_

_"I would like to have a few ideas in my head and run it through someone, but this is nothing formal."_

These interactions with their peers connect Participant 11 to other experts.

Finding the right literature is the starting point for preparing his experiments. For their last project, Participant 11 used an interesting paper from a different group; they examine the literature with their PI to discuss the findings.

## PLAN

Defining research goals is a collaborative effort accompanied by the PI and other seasoned scientists.

_"I will make a few slides explaining the idea."_

_"I am not going to plan too far; I am not going to spend a week planning because I do not know what I am going to end up with."_

Participant 11 has a clear goal in mind, but does not _"imagine it is [always] going to go well."_ They annotate literature before they execute experiments.

## ANALYZE

Analysis consists of trying multiple approaches and deciding what went well.

_"[The process] evolves, and you go back to the planning."_

Participant 11 conducts analysis using their notebook and computer software. _"The plotted graph is the final product."_ Participant 11 only re-uses raw data whenever they need to re-graph the plot.

## NOTE-TAKING

Before going to the lab, Participant 11 writes goals and targets on their notebook; they also copy protocols from an article they are trying to replicate. When using others' work, Participant 11 prefers to ask directly, _"but usually does get others' procedures until they are published."_ Post-its are temporary and can be easily lost, but lab notebooks can be brought out in those cases.

As Participant 11 conducts their experiments, they record their experimental steps, recording _"what [they] did to produce the sample […] in the lab notebook."_ They look at digitized data on the screen and then refer back to the notebook, since _"the notebook is the reference to everything [they] are doing."_

_"The lab notebook is mainly notes for [Participant 11]."_ Their note-taking convention is based on trial and error but has elements of others' note-taking style. _"You develop an intuition for what type of notes are useful or not."_ Participant 11 is trying to be more rigorous with their notes, since _"that is part of becoming a better scientist."_ They recreate or repeat experiments when details on the notebook are missing. Since the experiments can be easily rerun, it is not a big deal if some details are missing.

Participant 11 looks at the last pages to remind themselves on _"the stuff [they are] working on currently."_ They prefer to review presentations over looking at notebooks but occasionally refer to their notebooks to find more details. Supplementary pages in a manuscript are come from the details in their notebook and raw data. _"When you are writing a paper, you look at everything."_

## SAMPLE MANAGEMENT

_"Most of the naming identification comes when you are writing a paper."_

Participant 11 usually references a sample or batch from a specific page in the lab notebook. _"If you have the metadata in the name of the file, it is too long, and a random number does not help."_ They do not distinguish between a single sample and sample variations in a batch.

## DATA MANAGEMENT

Participant 11 _"always label [their] characterizations with date, initial, and page number"_ on their lab notebooks.

Raw files are not interesting; plotted and processed data is what's valuable. Participant 11 never deletes raw data, but does not look back at it once they have a successful characterization. _"I don't know why I would want to look back at it."_

## KNOWLEDGE TRANSFER

Knowledge is transferred in group discussions through mentoring and by doing things at the lab.

When Participant 11 presents, they target the content on the audience. The presentations contain _"the most complete work to date, including plotted and analyzed data"_ and Participant 11 _"would not direct other to raw files."_ They typically review processed data with others.

_"The raw data are not interesting, the plotted and processed data are what is good (…)_ d_o I care how many milligrams [my collaborator] put into the solution? Sometimes yes, sometimes I do not. I usually do not."_

Participant 11 does not know why metadata would be important to TRI, and _"does not get clear explanation of why."_

## USE OF E.L.N.

Participant 11 thinks ELNs can be useful but does not think that they are entirely well developed yet.

_"They are more trouble than they are beneficial. I have never really tried them, and I have not seen anybody in my group using them"_

Participant 11 is concerned with file flexibility and easy handwritten documentation. _"I just want ELNs to be user-accessible."_

When working on experiments, Participant 11 is continuously moving around the lab and expects an ELN to be portable.

## CONSIDERATIONS

1. **Quickly locate:** searchability would enable Participant 11 to see information across different experiments.

2. **Simple and fast:** It has to be fast and straightforward, _"just like the notebook."_

3. **Keep track of my place:** "_I know where I'm in the workflow. Because my workflow is essentially chronological._"

4. **Other's work:** Most of Participant 11's peers are working on different projects, so they _"don't need to see others' data."_

## DATA OWNERSHIP & ACCESS

_"Technically anyone in my group should be able to look at my lab notebook. They only time someone would want to repeat my products, is when I am not here."_

Beyond "my bad writing," they would not feel judged if other people see the content of their lab notebook. _"Nothing I am doing here is secretive."_

It is difficult to know what would be important for others when looking at his work _"did I record the things they want? Maybe not."_

# Participant 12

Participant 12 tries to understand how to design a new material that is better at what it does without having to do all the experiments. They do a lot of research, but they also mentor, train, and advise others in the lab.

Their autonomy is a trade-off: Participant 12 enjoys the autonomy, but may sometimes go too far in a fruitless direction.

*"For a senior member in lab, it's very loose as to exactly what you want to do. I am really interested in other people's projects and talking to them about how things are going and … how to best answer a question that's important."*

## Overview

### Relevant science is documented, art is not

There's a distinction, sometimes blurry, between "art" and "science". Science is what's strictly necessary for the result. Art is what can be done differently without affecting the result, such as setup details. The art may be helpful, but isn't important enough to share. Additionally, findings unrelated to your research question aren't report, even if they could be useful to others.

*"Sometimes [how you glue, cut, or flatten things] makes more of a difference than you would expect… We probably wouldn't describe … how we made this [setup] in a paper."*

### The lab notebook is only for the lab

The final destination of anything important is the computer, but the lab notebook contains lab work. Its content has transient value, mostly useful while in lab. This includes in-the-moment tracking, conditions to use in file naming, and problems or concerns.

*"I don't like to bring my notebooks outside of lab… I just have a general idea that lab is dirty… it means that I don't write a lot of things in my lab notebook if I'm sitting at my desk."*

### Think samples, not projects

Work starts with exploring samples and their properties. Only once you've found something new and interesting is there a project, which will turn into a manuscript. Samples exist before and independently of projects.

### Contingencies for what's next

Plans are short, mental decision trees. *Don't* plan too far ahead because outcomes are so uncertain. *Do* think ahead through what to do next so you can effectively make progress in lab whatever happens, whether because you're tired, distracted, or someone else is executing the plan.

**MOTIVATIONS** *"My job is about professional development… [academic positions] can be temporary and transient…my goal is to try and get to a point where I can get another job… I'd like to stay forever, but… that's not the plan for the position."*

**ON REPORTING** *"Nobody wants to report bad data. Nobody wants to report things that aren't active. I think that's a failing in our field."*

## EXPLORE

Participant 12 goes to conferences and thinks about questions people are asking. Once they have a question, they read papers to see if it's been answered. They may re-analyze others' data for their question to see if there can gain interesting insights without having to do new experiments. The experimental section is useful, but many people leave out details (e.g., setup) that make the process easier. If Participant 12 knows the author, they email them for more details; sometimes, readers of their publications email Participant 12. The world gets smaller over time.

## PLAN

Participant 12 usually plans at their desk while looking at data, thinking through their next steps in comfort. The plan is a mental decision tree. Participant 12 only writes it down if others are executing it. Otherwise, they would have to cross many things out. Since their lab notebook doesn't leave lab due to contamination concerns, Participant 12 writes to-do lists on sticky notes, such as data to collect, that they then bring to lab. These notes are so that Participant 12 collects the intended data, even if they are tired or distracted; the notes have no value after Participant 12 has collected that data. Days without a formal plan don't work out so well.

## ANALYZE

Participant 12 occasionally takes pictures of their lab notebook for later reference at their computer, though not often. Usually, the data files and their filenames are enough.

Every technique has a different type of analysis that may involve Excel, proprietary software, and working through by hand. Participant 12 wishes more was automated, but their coding skills are not sufficient to automate the analysis.

Participant 12 sometimes analyzes data that they didn't collect and don't have all the details for, so they ask others for clarification.

## NOTE-TAKING

Participant 12's notebook contains observational notes, mostly about data being nullified (did something wrong) or concerning (something is off). Some information is only important in-the-moment, such as tracking progress or sample holder location. Participant 12 has a system for cross-referencing their notes and their data, though almost everything important is on the computer.

Sometimes timing information is only in the lab notebook, if anywhere. It's often not important, but it's important for their current research that involves an unstable material. A groupmate working on the same project was struggling with reproducibility. Participant 12 and the collaborator eventually had someone apply machine learning on their information and find that samples made on a specific day worked because it corresponded to a cleaning schedule. This took months to figure out.

*"My notebooks are very useless to other people, I apologize for that… I think of it as my key for data on my computer".*

## SAMPLE MANAGEMENT

Not covered explicitly.

They do have some samples that they keep and sometimes revisit later.

## DATA MANAGEMENT

Anything important from the notebook goes in the filename, including date, lab notebook # and page, title, sample, experimental conditions. All information is saved, even exploratory and bad data.

Participant 12 organize folders by date and technique. Participant 12 works with collaborators and have data on both a shared drive and other collaborative platforms. They cannot find the other person's data without asking the person who collected the data.

## USE OF E.L.N.

Participant 12 has never used a formal ELN.

For highly collaborative experiments, only collaborative cloud platform is used (no physical lab notebook). The environment is more like an office than the lab. Execution is highly collaborative. Participant 12 documents a decision tree so others know the plan and can make decisions when they are unavailable. Detailed notes are taken, as analysis takes a long time and it can help distinguish between instrumentation errors and other problems.

## CONSIDERATIONS

1. **Enter after, not before:** filling out the workflow before executing it doesn't make sense, as Participant 12 "*would need to edit constantly*".

2. **I've tried this before:** with Excel. Participant 12 used for a few weeks and then gave up. Participant 12 has enough interest to try something new, but so far, has found nothing useful enough to keep going.

3. **Organize by sample:** this is how Participant 12 thinks about their own work. They want an easier way to find data. Right now, they look for the date on which something was made (in the notebook), and cross reference with a folder on the computer for data.

## KNOWLEDGE TRANSFER

Group presentations can be given at any stage of the process. For example, when exploring, Participant 12 may present what they intend to work on and any preliminary experiment(s). When others present, Participant 12 follows up with them to build relationships and provide feedback to their work. Manuscripts can leave out information that is important to Participant 12, such as setup information or findings unrelated to the research question.

*"This material is not new in the literature, so it was really frustrating for me that no one had mentioned it wasn't stable… I present posters on it and got some really delightful feedback from a professor like, 'Why did you keep going on this project? It's clear that this material isn't stable.'"*

## DATA OWNERSHIP & ACCESS

Participant 12 and their group should have access to everything. However, some collaborators do things they don't do (e.g., synthesis). In these cases, Participant 12 does not want to see synthetic data, because they would have to ask too many questions to fully understand the data.

Participant 12 has made raw and processed data publicly available due to funding agency requirements, especially government. They like this, since this means that they can access others' data. But it's also complicated, because putting up raw data doesn't mean people will understand it.

*"I feel less worried about… being scooped… I like to think about it more as a sharing of knowledge that will progress humanity in a very naïve way."*

# Participant 13

Participant 13 is in multiple groups. PIs give Participant 13 feedback in group and individual meetings. Participant 13 interacts with other members of the group to learn new techniques. Each project has 1-2 outside collaborators. If you want to publish in a high impact journal, you need some advanced characterization instruments which they don't have at the university.

## Overview

### My organization is good enough for me…

…but not for future group members. Participant 13 has needed to go through work published by previous members of the group to find trends, which involved a lengthy search through others' files. *"Right now, I haven't had to give [data] to other people. But after I leave, there might be someone who wants to replicate my data and it would be hard for them."* Although Participant 13 has some challenges searching through their own information, it's not that big of a deal for them. *"I know the data is there, but whenever I want to find it, I don't know where it is…. It [mostly takes] within 10 minutes, but still less efficient."*

### Sample naming is hard

Sample names are an index across different kinds of info, but for Participant 13, this index does not always uniquely identify links and can be broken. For example, Participant 13 may have synthesized "[composition]-[material]-[##]" several times. *"When I want to publish, I will look for the optimized synthetic condition. Even for me, it takes some time to find it."* Participant 13 may need to experimentally confirm which synthesis procedure was used in preparation for publishing. This working sample name is replaced in publications, making it hard to later find data associated with a published figure.

### Presentations don't include everything that matters

In meetings, *"most of the time, people won't care about the synthetic conditions, only the performance."* For this reason, synthesis info only exists in the lab notebook until manuscript drafting.

### Some raw data is needed by others

Rawness of data is a spectrum. When Participant 13 wants "raw" data, it is really "semi-processed" data he can re-analyze or re-plot, such as the XY data points for a graph. They don't need anything more raw, but just the final, static form (e.g., graph) is not enough.

**ON PURSUING NEW IDEAS** *"Many times, this project is not working out, so it will stop halfway."*

**ON THE PROTOTYPE** *"I look forward to the software… it never hurts. If it doesn't work well, we can just abandon it! If it's really great, we'll change our style and use it."*

## EXPLORE

*"If I have an idea, I'll put it in my [non-lab] notebook and try it out. There are many ideas."* Some come from attending seminars and conferences. An idea might be a drawing of a molecule to try out someday.

## PLAN

Participant 13 searches for a procedure for something similar to their target goal. I look at published conditions and think about what makes sense for Participant 13's target. If a setup is very specific or special, they will sketch it into the notebook. They might check with PIs for feedback. Once this is done, Participant 13 tries it in the lab.

Because things usually don't work, they typically revisit the plan after analysis.

## ANALYZE

All the instruments Participant 13 uses are connected to a computer. They export the data to their own computer for analysis. Participant 13 goes through multiple cycles of execute and analyze until they have enough data. *"Enough depends on how much time you have."* If the goal is to publish in a high impact journal, *"it's never going to be enough."* For example, when writing a manuscript, Participant 13 realized that their characterization was not enough for a high impact journal, so they collaborated with others for more advanced characterization. Since participant 13 does not know how to process the raw data, they request an Origin file from them, not the raw instrument files used to create the Origin file. *"It's not the final figure, but it's data you can process by yourself."*

## NOTE-TAKING

The lab notebook *"mostly contains the procedure of the synthesis [Participant 13 is] conducting,"* which are protocols *"that cannot be efficiently recorded by computer."* Participant 13 finds it easier and faster to draw a reaction on paper than to do on a computer. Anything important during an experiment will be written in the lab notebook. Participant 13 uses loose paper for notes that are not considered important. *"The lab notebook doesn't contain any experimental data."* All data is in Participant 13's laptop. If the data shows a problem, Participant 13 might record "not working," in their notebook.

They refer back to their notebook when writing a manuscript, to find their synthesis conditions or to repeat an experiment with a larger batch. If the setup was very different than the literature, they may have a crude drawing in their lab notebook which can be redrawn in Photoshop. *"If I lost the lab notebook, that would be a serious problem… before I publish, the only copy [of synthesis conditions] is here."*

## SAMPLE MANAGEMENT

*"When I explore this synthetic condition, I will write the label as the date of the experiment… Once I find the optimized synthetic conditions… I will name that material."* But when writing a manuscript, sometimes this name won't uniquely identify the recipe, so it needs to be experimentally confirmed. *"That's why it takes so long to publish something, because you need to confirm a lot of data. When you run this test, you probably didn't expect this test to be useful, so you didn't label it very precisely."*

## DATA MANAGEMENT

Whenever Participant 13 wants to try something new, they create a folder on their desktop with the current data. Within that folder, they will create a new folder for each instrument, as well as new folders for figures, references, and manuscripts.

*"I need a copy [of data] on my computer, for safety. In the shared folder, someone might accidentally delete the folder. You'd lose all your data!"*

## USE OF E.L.N.

Never used an ELN—not specifically discussed.

## CONSIDERATIONS

1. **Quickly locate my + others' info:** I've had to look through past group member's files to find data. *"This really took a long time."* Even with my own files, *"there's so much data in different places"* that it can me 10 minutes to find what I'm looking for. Windows search isn't even helpful, *"if you have a dash between the names, it can't find it."*

2. **Batch upload:** When I perform XRD, I may have a batch of 20 in a day. Right now, I just copy to these files to a flash drive. I don't want something much more time consuming.

3. **Use my data to help me:** Although drawing a chemical formula would take me more time on the computer than by-hand, I would do it if it helped me in other ways. For example, if it could perform reaction amount calculations for me, I might consider it.

## KNOWLEDGE TRANSFER

*"We're a small group, we have time for everyone to present their work… Personally, I do not spend a lot of time preparing for these meetings. Whenever I collect data, I process it and make a plot. Everything is ready. Whenever I need to talk about it, I have figures in hand, so it's easier to make the slides."*

Participant 13 always has results within a month to share with PIs and group members and to give a status update. They share once there's enough data to have an interpretation.

## DATA OWNERSHIP & ACCESS

*"If TRI wants raw data, I can definitely share"* if it's just a matter of uploading data, but Participant 13 doubts that it will be useful. Without enough context, some raw data cannot be understood. *"There's nothing we need to hide from TRI."*

Exploratory information and planning is for Participant 13 and perhaps their PI. Meetings with the group do not affect Participant 13's research plan, though their feedback may influence the direction of Participant 13's work once the work is under way and there are results to discuss.